\documentclass[aps,prd,preprint]{revtex4-1}
\usepackage{graphicx}
\usepackage{amsmath,amssymb, graphics, setspace,longtable,array,color}
\usepackage[export]{adjustbox}
\usepackage{commath}
\usepackage[export]{adjustbox}
\usepackage{courier,float}
\usepackage{enumitem}
\usepackage{siunitx}
\usepackage{relsize}
\usepackage{multirow}
\usepackage{todonotes}
\usepackage{ulem}
\usepackage{cancel}

\newcommand{\mathsym}[1]{{}}
\newcommand{\unicode}[1]{{}}

\makeatletter
\def\p@subsection{}
\def\p@subsubsection{}
\def\p@paragraph{}
\def\p@subparagraph{}
\makeatother

\makeatletter
\def\l@section{\@dottedtocline{1}{1em}{2em}}
\def\l@subsection{\@dottedtocline{2}{1.5 em}{2em}}
\def\l@subsubsection{\@dottedtocline{3}{2em}{3em}}
\def\l@paragraph{\@dottedtocline{4}{2.5em}{4em}}
\def\l@subparagraph{\normalfont \@dottedtocline{5}{3.5 em}{4 em}}
\makeatother

\usepackage{fancyhdr}

\begin{document}

\title{Time Delay Interferometry combinations as instrument noise monitors for LISA}

\author{Martina~Muratore}\email{contact: martina.muratore@unitn.it}\affiliation{\addressi}
\author{Daniele~Vetrugno}\affiliation{\addressi}
\author{Stefano~Vitale}\email{contact: stefano.vitale@unitn.it}\affiliation{\addressi}
\author{Olaf~Hartwig}\affiliation{\addres}

\def\addres{Max-Planck-Institut für Gravitationsphysik (Albert-Einstein-Institut), Callinstraße 38, 30167 Hannover, Germany}
\def\addressi{Dipartimento di Fisica, Universita di Trento and Trento Institute for 
Fundamental Physics and Application / INFN, 38123 Povo, Trento, Italy}
\date{May 16, 2020}

\begin{abstract}

The LISA mission will likely be a signal dominated detector, such that one challenge is the separation of the different astrophysical sources, and to distinguish between them and the instrumental noise.  One of the goals of LISA is to probe the early universe by detecting stochastic gravitational wave (GW) backgrounds. As correlation with other detectors is uncertain for LISA, discriminating such a GW background from the instrumental noise requires a good estimate of the latter. To this purpose we have revisited Time Delay Interferometry (TDI) to look for new TDI signal combinations that fulfill the laser frequency noise suppression requirements. We illustrate that it is possible to do a linear combination of these TDI channels to find special null-combinations that suppress GWs and mainly carry information about instrumental noise. We find that there exist many null-combinations that show different sensitivities to GWs, some of which seem more suitable than the traditional T combination for estimating test-mass (TM) acceleration noise. In an idealised LISA configuration, they are all sensitive to a particular linear combination of the six TMs acceleration, similar to a rigid rotation of the LISA triangle. \\
In the following article, we illustrate what are the noise properties that can be extracted by monitoring these interferometry signals and discuss the implication of these findings for the detection of stochastic GW backgrounds.
\end{abstract}

\maketitle

\section{Introduction}
We have recently revisited the TDI combinations that suppress laser noise in LISA. In \cite{Muratore_2020} are reported 174 combinations of 16-links and 12 of 12 links. By reviewing ref. \cite{Muratore_2020} and applying the methodology explained there, we realised to have skipped 24 combinations of 14-links that also fullfill the frequency noise suppression requirements \cite{Muratore:2021phd,Hartwig:2021phd}. Thus, the total TDI combinations are 210. These combinations could be reduce to a subset of 28 of 16-links, 3 of 14-links and 3 of 12-links, if one considers as equal combinations that also differ for time reversal symmetry in addition to the satellites permutations symmetries considered in \cite{Muratore_2020}. We consider these 34 combinations as the core subset of the 210 total combinations. \\ The LISA detector is expected to be signal dominated, thus to measure the instrumental noise we look for TDI combinations that have suppressed sensitivity to gravitational wave (GW)  signals but still carry some information on the instrumental noise \footnote{Note that other space missions are being planed such as the Chinese Taiji mission. Thus, in principle we might use correlation with other space-based GW detectors to detect the stochastic GW background. However, differently from LISA that has been selected to be ESA's third large-class mission \cite{amaroseoane2017laser}, the Taiji detector is not at the same stage of development. Thus in this paper we will perform our studies considering a single space-based LISA-like detector}. We find that a good proxy for such a property is for the combinations to be perfectly insensitive to GW when the arm lengths are equal and constant and the wave is propagating normally to such a perfect idealised LISA configuration. In this ideal condition, the signal out of any TDI combination becomes a linear combination of a finite set of single link signals with delays also chosen from a finite set. This allows to explore the space of all linear combinations of all possible TDI combinations, and to identify its null space, i.e. the set of all combinations that gives zero response to GW.  \\ We find that the rank of such null space is just 7 and that the 7 fundamental combinations all give a, differently delayed,  measurement of the same combination of TM accelerations. This combination has a symmetry that resembles the rigid rotation of the constellation, and is different from that which enters in GW sensitive TDI combinations as, for instance, the second generation Michelson X.  \\

The article is divided in five main sections. In the first section we use linear algebra to determine which are the TDI combinations that are insensitive to GW signals in case of a perfect orthogonal source considering a static LISA with equal arm-length. In the second section we use the method of single value decomposition to retrieve the instrumental noise of these combinations. In the third section, we give an estimate of the behaviour of these null-channels in case the LISA constellation undergoes small amplitude static distortions. Moreover, we compute the sensitivity both for a perfect LISA triangle and a non perfectly equilateral triangle, and we also validate our model of the TM acceleration noise with the outcome of numerical simulations performed using the python tools LISANode \cite{Bayle:2019phd} and PyTDI. Here, we show that some combinations deviate significantly from the simplified case when taking inequalities in the arms into account. For example, the traditional T null channel appears to have the same GW sensitivity as X in this more realistic scenario. The last section reports our conclusion and future perspective regarding the leverage of these null channels to calibrate the instrument during operations and we discuss the implication of these findings for the detection of stochastic GW backgrounds. 

\section{\label{NULL} TDI Combinations with reduced sensitivity to gravitational waves}

The full list of TDI combinations up to a length of 16 links has been revisited with respect to  \cite{Muratore_2020}. The lists of all 28 core combinations of 16 links, plus the 3 combinations of 12 links and the 3 of 14 links are shown in Tables \ref{T16l}, \ref{T14l} and \ref{T12l}. 		\\
The combinations within this minimum set fall in a few simple categories, marked by: the number of emission/measurements event pairs, which ranges from 1 to 6, the number of arms involved, 2 or 3 and finally the number of inter-satellite links, 4 or 6, out of the six possible ones ($1\to2,\;2\to1,\;1\to3,\;3\to1,\;2\to3,\;\text{and}\;3\to2$) that are involved in the sequence.  As an example, the standard TDI X, that in our notation is $C^{16}_1$, shows up on the first line of Table \ref{T16l}. It only involves 2 arms, 4 inter-SC links and 1 measurement event. \\
The TDIs $C^{12}_{3}$, $C^{14}_{3}$, $C^{16}_{26}$, $C^{16}_{27}$ have the properties of being fully symmetric in the idealistic scenario of equal and constant arm-length. In fact, they represent second generation versions of the first generation fully symmetric Sagnac combination known from the literature \cite{Tinto2020}. Note, however, that they are different from the second generation versions reported there.\\
We will mostly focus on the fully symmetric Sagnac combination with the minimum number of links that is the $C^{12}_{3}$ to which we simply refer by $\zeta$ in the following.\\

See Supplemental Material at [URL] for the full set of 210 TDI combinations and the Appendix \ref{fullTDI} for a complete explanation of how to get the full set of TDI combinations from the core.
 \begin{table}[H]
\begin{center}
\resizebox{1.01\textwidth}{.1\textheight}{
\tiny
\begin{tabular}{|c|c|c|c|c||c|c|c|c|c|}
\hline
 Number of & Combination & Number of & Number of  & Number of  &   Number of  & Combination & Number of  & Number of  & Number of    \\ 
  combination &  & measurements & arms & inter-SC-links&   combination &  & measurements & arms & inter-SC-links \\ 
  \hline
$ \mathcal{C}_1^{16}$ & 1$\rightarrow $2$\rightarrow $1$\rightarrow $3$\rightarrow $1$\rightarrow $3$\rightarrow $1$\rightarrow $2$\rightarrow $1$\leftarrow $3$\leftarrow $1$\leftarrow $2$\leftarrow $1$\leftarrow $2$\leftarrow $1$\leftarrow
   $3$\leftarrow $1 & 1 & 2 & 4 & $\mathcal{C}_{15}^{16}$ & 1$\rightarrow $2$\rightarrow $1$\rightarrow $3$\rightarrow $2$\leftarrow $1$\leftarrow $2$\leftarrow $3$\rightarrow $1$\rightarrow $2$\rightarrow $1$\leftarrow $3$\leftarrow $2$\leftarrow
   $1$\leftarrow $2$\rightarrow $3$\leftarrow $1 & 3 & 3 & 6 \\
$ \mathcal{C}_2^{16}$ & 1$\rightarrow $2$\rightarrow $1$\rightarrow $3$\rightarrow $2$\rightarrow $3$\rightarrow $1$\rightarrow $2$\rightarrow $1$\leftarrow $3$\leftarrow $2$\leftarrow $1$\leftarrow $2$\leftarrow $1$\leftarrow $2$\leftarrow
   $3$\leftarrow $1 & 1 & 3 & 6 & $\mathcal{C}_{16}^{16}$ & 1$\rightarrow $2$\rightarrow $1$\rightarrow $3$\leftarrow $2$\leftarrow $1$\leftarrow $2$\rightarrow $3$\rightarrow $2$\leftarrow $1$\leftarrow $2$\leftarrow $3$\rightarrow $1$\rightarrow
   $2$\rightarrow $1$\leftarrow $3$\leftarrow $1 & 3 & 3 & 6 \\
$ \mathcal{C}_3^{16}$ & 1$\rightarrow $2$\rightarrow $3$\rightarrow $1$\rightarrow $2$\rightarrow $1$\rightarrow $3$\rightarrow $2$\rightarrow $1$\leftarrow $3$\leftarrow $2$\leftarrow $1$\leftarrow $2$\leftarrow $1$\leftarrow $2$\leftarrow
   $3$\leftarrow $1 & 1 & 3 & 6 &$ \mathcal{C}_{17}^{16}$ & 1$\rightarrow $2$\rightarrow $1$\rightarrow $3$\rightarrow $2$\leftarrow $1$\leftarrow $3$\leftarrow $2$\leftarrow $1$\rightarrow $3$\rightarrow $1$\rightarrow $2$\leftarrow $3$\leftarrow
   $1$\leftarrow $2$\rightarrow $3$\leftarrow $1 & 3 & 3 & 6 \\
 $\mathcal{C}_4^{16} $& \text{1$\rightarrow $2$\rightarrow $1$\rightarrow $2$\rightarrow $1$\leftarrow $3$\leftarrow $1$\leftarrow $2$\leftarrow $1$\rightarrow $3$\rightarrow $1$\rightarrow $3$\rightarrow $1$\leftarrow $2$\leftarrow $1$\leftarrow
   $3$\leftarrow $1} & 2 & 2 & 4 &$\mathcal{C}_{18}^{16} $& \text{1$\rightarrow $2$\rightarrow $1$\rightarrow $3$\leftarrow $2$\leftarrow $1$\rightarrow $3$\rightarrow $2$\rightarrow $1$\leftarrow $3$\leftarrow $1$\leftarrow $2$\rightarrow $3$\rightarrow
   $1$\leftarrow $2$\leftarrow $3$\leftarrow $1} & 3 & 3 & 6 \\
$\mathcal{C}_5^{16}$ & \text{1$\rightarrow $2$\rightarrow $1$\rightarrow $3$\rightarrow $1$\rightarrow $2$\rightarrow $1$\leftarrow $3$\leftarrow $1$\leftarrow $2$\leftarrow $1$\rightarrow $3$\rightarrow $1$\leftarrow $2$\leftarrow $1$\leftarrow
   $3$\leftarrow $1} & 2 & 2 & 4 & $\mathcal{C}_{19}^{16}$ & \text{1$\rightarrow $2$\rightarrow $3$\rightarrow $2$\rightarrow $3$\rightarrow $2$\rightarrow $1$\leftarrow $3$\leftarrow $2$\leftarrow $1$\rightarrow $3$\leftarrow $2$\leftarrow $3$\rightarrow
   $1$\leftarrow $2$\leftarrow $3$\leftarrow $1} & 3 & 3 & 6 \\
$\mathcal{C}_6^{16}$ & \text{1$\rightarrow $2$\rightarrow $1$\rightarrow $3$\rightarrow $2$\rightarrow $1$\rightarrow $2$\leftarrow $3$\leftarrow $1$\leftarrow $2$\leftarrow $1$\rightarrow $3$\rightarrow $2$\leftarrow $1$\leftarrow $2$\leftarrow
   $3$\leftarrow $1} & 2 & 3 & 4 & $\mathcal{C}_{20}^{16}$ & \text{1$\rightarrow $2$\rightarrow $3$\rightarrow $2$\rightarrow $3$\rightarrow $2$\rightarrow $1$\leftarrow $3$\leftarrow $2$\leftarrow $3$\rightarrow $1$\leftarrow $2$\leftarrow $1$\rightarrow
   $3$\leftarrow $2$\leftarrow $3$\leftarrow $1} & 3 & 3 & 6 \\
$ \mathcal{C}_7^{16}$ & \text{1$\rightarrow $2$\rightarrow $3$\rightarrow $1$\rightarrow $2$\rightarrow $3$\leftarrow $1$\leftarrow $3$\leftarrow $2$\leftarrow $1$\rightarrow $3$\rightarrow $1$\rightarrow $3$\leftarrow $2$\leftarrow $1$\leftarrow
   $3$\leftarrow $1} & 2 & 3 & 4 & $\mathcal{C}_{21}^{16}$ & \text{1$\rightarrow $2$\rightarrow $1$\rightarrow $2$\rightarrow $3$\leftarrow $1$\leftarrow $2$\leftarrow $1$\rightarrow $3$\leftarrow $2$\rightarrow $1$\rightarrow $3$\leftarrow $2$\leftarrow
   $1$\leftarrow $2$\rightarrow $3$\leftarrow $1} & 4 & 3 & 4 \\
$ \mathcal{C}_8^{16}$ & \text{1$\rightarrow $2$\rightarrow $3$\rightarrow $1$\rightarrow $3$\rightarrow $1$\rightarrow $2$\rightarrow $3$\leftarrow $1$\leftarrow $3$\leftarrow $2$\leftarrow $1$\rightarrow $3$\leftarrow $2$\leftarrow $1$\leftarrow
   $3$\leftarrow $1} & 2 & 3 & 4 &$ \mathcal{C}_{22}^{16}$ & \text{1$\rightarrow $2$\rightarrow $1$\rightarrow $3$\leftarrow $2$\leftarrow $1$\leftarrow $2$\rightarrow $3$\leftarrow $1$\leftarrow $2$\leftarrow $1$\rightarrow $3$\leftarrow $2$\rightarrow
   $1$\rightarrow $2$\rightarrow $3$\leftarrow $1} & 4 & 3 & 4 \\
$ \mathcal{C}_9^{16}$ & \text{1$\rightarrow $2$\rightarrow $1$\rightarrow $2$\rightarrow $1$\leftarrow $3$\leftarrow $2$\leftarrow $1$\leftarrow $2$\rightarrow $3$\rightarrow $1$\rightarrow $3$\rightarrow $2$\leftarrow $1$\leftarrow $2$\leftarrow
   $3$\leftarrow $1} & 2 & 3 & 6 & $\mathcal{C}_{23}^{16}$ & \text{1$\rightarrow $2$\rightarrow $1$\rightarrow $2$\rightarrow $1$\leftarrow $3$\rightarrow $2$\leftarrow $1$\leftarrow $2$\leftarrow $3$\rightarrow $1$\rightarrow $3$\leftarrow $2$\leftarrow
   $1$\leftarrow $2$\rightarrow $3$\leftarrow $1} & 4 & 3 & 6 \\
 $\mathcal{C}_{10}^{16}$ & \text{1$\rightarrow $2$\rightarrow $1$\rightarrow $3$\rightarrow $1$\rightarrow $2$\rightarrow $1$\leftarrow $3$\leftarrow $2$\leftarrow $1$\leftarrow $2$\rightarrow $3$\rightarrow $2$\leftarrow $1$\leftarrow $2$\leftarrow
   $3$\leftarrow $1} & 2 & 3 & 6 & $\mathcal{C}_{24}^{16}$ & \text{1$\rightarrow $2$\rightarrow $3$\rightarrow $1$\leftarrow $2$\leftarrow $1$\rightarrow $3$\leftarrow $2$\leftarrow $1$\rightarrow $3$\rightarrow $2$\rightarrow $1$\leftarrow $3$\leftarrow
   $1$\rightarrow $2$\leftarrow $3$\leftarrow $1} & 4 & 3 & 6 \\
$ \mathcal{C}_{11}^{16} $& \text{1$\rightarrow $2$\rightarrow $3$\rightarrow $1$\rightarrow $2$\rightarrow $3$\rightarrow $2$\rightarrow $1$\leftarrow $3$\leftarrow $2$\leftarrow $1$\rightarrow $3$\leftarrow $2$\leftarrow $1$\leftarrow $2$\leftarrow
   $3$\leftarrow $1} & 2 & 3 & 6 & $\mathcal{C}_{25}^{16}$ & \text{1$\rightarrow $2$\rightarrow $3$\rightarrow $2$\rightarrow $1$\leftarrow $3$\leftarrow $2$\leftarrow $3$\rightarrow $1$\leftarrow $2$\rightarrow $3$\rightarrow $2$\leftarrow $1$\rightarrow
   $3$\leftarrow $2$\leftarrow $3$\leftarrow $1} & 4 & 3 & 6 \\
 $\mathcal{C}_{12}^{16}$ & \text{1$\rightarrow $2$\rightarrow $3$\rightarrow $1$\rightarrow $3$\rightarrow $2$\rightarrow $1$\leftarrow $3$\leftarrow $1$\leftarrow $2$\leftarrow $3$\rightarrow $1$\rightarrow $3$\leftarrow $2$\leftarrow $1$\leftarrow
   $3$\leftarrow $1} & 2 & 3 & 6 & $\mathcal{C}_{26}^{16}$ & \text{1$\rightarrow $2$\rightarrow $1$\rightarrow $2$\rightarrow $1$\leftarrow $3$\rightarrow $2$\leftarrow $1$\rightarrow $3$\leftarrow $2$\leftarrow $1$\leftarrow $2$\leftarrow $3$\rightarrow
   $1$\leftarrow $2$\rightarrow $3$\leftarrow $1} & 5 & 3 & 6 \\
 $\mathcal{C}_{13}^{16}$ & \text{1$\rightarrow $2$\rightarrow $1$\rightarrow $2$\rightarrow $1$\leftarrow $3$\leftarrow $2$\leftarrow $1$\rightarrow $3$\rightarrow $2$\leftarrow $1$\leftarrow $2$\rightarrow $3$\rightarrow $1$\leftarrow $2$\leftarrow
   $3$\leftarrow $1} & 3 & 3 & 6 & $\mathcal{C}_{27}^{16}$ & \text{1$\rightarrow $2$\rightarrow $1$\rightarrow $3$\rightarrow $2$\leftarrow $1$\rightarrow $3$\leftarrow $2$\rightarrow $1$\leftarrow $3$\leftarrow $1$\leftarrow $2$\leftarrow $3$\rightarrow
   $1$\leftarrow $2$\rightarrow $3$\leftarrow $1} & 5 & 3 & 6 \\
 $\mathcal{C}_{14}^{16} $& \text{1$\rightarrow $2$\rightarrow $1$\rightarrow $3$\rightarrow $1$\rightarrow $2$\rightarrow $1$\leftarrow $3$\rightarrow $2$\leftarrow $1$\leftarrow $2$\leftarrow $3$\leftarrow $2$\leftarrow $1$\leftarrow $2$\rightarrow
   $3$\leftarrow $1} & 3 & 3 & 6 & $\mathcal{C}_{28}^{16}$ & 1$\rightarrow $2$\rightarrow $1$\leftarrow $3$\leftarrow $2$\rightarrow $1$\leftarrow $3$\rightarrow $2$\leftarrow $1$\rightarrow $3$\rightarrow $1$\leftarrow $2$\leftarrow $3$\rightarrow
   $1$\leftarrow $2$\rightarrow $3$\leftarrow $1 & 6 & 3 & 6 \\  \hline
 \end{tabular}}
\caption{\footnotesize{16-links TDI  combinations that suppress laser noise. $C_{i}^{16}$ is the sequence number. Each combinations is represented with numbers and arrows: numbers within each sequence indicate the satellite on which each event takes place. First and last event coincide. Arrows indicate the direction of the link connecting adjoining events. Events at the start of two arrows represent simultaneous emission of two beams. Events at the end of two arrows represent measurements. Events between arrows going in the same direction indicate that the spacecraft (SC) transponders/relays the beam to the next that follows. The first  is always an emission event. The last three columns indicates the number of measurements events, the number of arms and number of inter-SC links involved in the sequence, respectively.} \label{T16l}}
\end{center}
\end{table}
\begin{table}[H]
\begin{center}
\resizebox{1.01\textwidth}{.035\textheight}{
\begin{tabular}{|c|c|c|c|c||c|c|c|c|c|}
\hline
 Number of & Combination & Number of & Number of  & Number of  &   Number of  & Combination & Number of  & Number of  & Number of   \\ 
  combination &  & measurements & arms & inter-SC-links&   combination &  & measurements & arms & inter-SC-links \\ \hline
$ \mathcal{C}_1^{14}$ & \text{1$\rightarrow $2$\rightarrow $1$\rightarrow $3$\rightarrow $2$\rightarrow $1$\leftarrow $3$\leftarrow $2$\leftarrow $1$\leftarrow $2$\rightarrow $3$\rightarrow $1$\leftarrow $2$\leftarrow $3$\leftarrow $1} & 2 & 3 & 6 &
  $ \mathcal{C}_3^{14}$ & \text{1$\rightarrow $2$\rightarrow $1$\rightarrow $3$\leftarrow $2$\rightarrow $1$\leftarrow $3$\rightarrow $2$\leftarrow $1$\leftarrow $2$\leftarrow $3$\rightarrow $1$\leftarrow $2$\rightarrow $3$\leftarrow $1} & 5 & 3 & 6 \\
$ \mathcal{C}_2^{14}$ & \text{1$\rightarrow $2$\rightarrow $1$\rightarrow $3$\leftarrow $2$\leftarrow $1$\leftarrow $3$\rightarrow $2$\leftarrow $1$\leftarrow $2$\leftarrow $3$\rightarrow $1$\rightarrow $2$\rightarrow $3$\leftarrow $1} & 3 & 3 & 6 &
 &  &  &  &  \\
  \hline
\end{tabular}}
\end{center}
\caption{\footnotesize{14-link TDI  combinations that suppress laser noise. Definitions are the same as those for TABLE \ref{T16l}}}
\label{T14l}
\end{table}
\begin{table}[H]
\begin{center}
\resizebox{1.01\textwidth}{.035\textheight}{
\begin{tabular}{|c|c|c|c|c||c|c|c|c|c|}
\hline
 Number of & Combination & Number of & Number of  & Number of  &   Number of  & Combination & Number of  & Number of  & Number of    \\ 
  combination &  & measurements & arms & inter-SC-links&   combination &  & measurements & arms & inter-SC-links \\ \hline
$ \mathcal{C}_1^{12}$ & \text{1$\rightarrow $2$\rightarrow $3$\rightarrow $1$\rightarrow $3$\rightarrow $2$\rightarrow $1$\leftarrow $3$\leftarrow $2$\leftarrow $1$\leftarrow $2$\leftarrow $3$\leftarrow $1} & 1 & 3 & 6 & $\mathcal{C}_3^{12} $&
   \text{1$\rightarrow $2$\rightarrow $1$\leftarrow $3$\rightarrow $2$\leftarrow $1$\rightarrow $3$\leftarrow $2$\leftarrow $3$\rightarrow $1$\leftarrow $2$\rightarrow $3$\leftarrow $1} & 5 & 3 & 6 \\
$ \mathcal{C}_2^{12}$ & \text{1$\rightarrow $2$\rightarrow $3$\rightarrow $2$\rightarrow $1$\leftarrow $3$\leftarrow $2$\leftarrow $1$\rightarrow $3$\rightarrow $1$\leftarrow $2$\leftarrow $3$\leftarrow $1} & 2 & 3 & 6 &  &  &  &  &  \\
  \hline
\end{tabular}}
\caption{\footnotesize{12-link TDI  combinations that suppress laser noise. Definitions are the same as those for  TABLE \ref{T16l}.}}
\label{T12l}\end{center}
\end{table}

\begin{figure}
\centering
\includegraphics[width=0.6\linewidth]{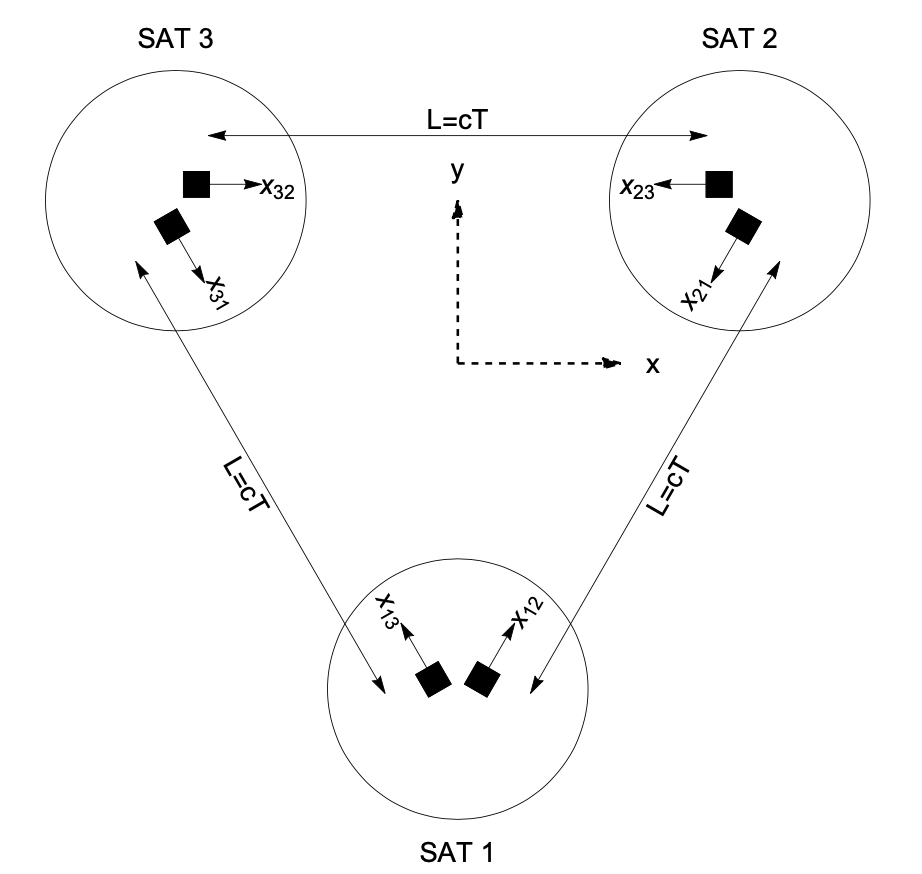}
\caption{Schematics of the idealised LISA geometry. Three satellites (indicated as SAT), shown as circles, include two test-masses, shown as black squares, each. The satellites are at the vertexes of an equilateral triangle in the x-y plane, with orientation relative to the axes as shown in the picture. The figure also shows the numbering convention for the satellites, and the orientation chosen as positive, for the displacement $x_{j,i}$ of the TM in satellite $j$ at the end of the link from satellite $i$.}
\label{LISA}
\end{figure}

In order to look for TDI combinations that are insensitive to GW signals, we calculate their sensitivity to GW  coming orthogonal to LISA, in the idealised equilateral, constant arm-length configuration  (see Fig. \ref{LISA}). Indeed, to do a quick search one can focus on frequencies for which the maximum delay among the various link signals, 8$\tau$, has negligible impact, that is, for which  $2\pi f\times 8\tau \ll 1$. For $c \tau = L = \SI{2.5}{\giga\meter}$ this suggests frequencies below \SI{20}{\milli\hertz}.\\
 A TDI combinations that gives zero signal in this simplified case is expected to suppress the signal in realistic situation. An example of these TDI combinations that we are looking for is the combination known as T, given for example in \cite{PhysRevD.89.022001}.\\ 
Regarding the instrumental noises, we consider only the effect of acceleration noise, as this is going to be the dominant noise source of the LISA instrument below a few \si{\milli\hertz} \footnote{ The acceleration noise is any local disturbance acting on the test-mass (TM) that depends on many physical effects which locally perturbs the LISA TM geodesic motion inside the Gravitational Reference Sensor. It can be measured as an acceleration of the TM with respect to a local inertial reference frame, and has been demonstrated by LISA Pathfinder to be within the mission requirements \cite{article}.}. \\

We can summarize the LISA measurements principle by considering how the GW signal enters into a single inter-spacecraft link. One of the satellites, called the emitter,  sends a beam to a second one, the receiver, which measures the frequency or phase of the received beam. GWs are detected as a modulation of the frequency or phase of the received beam. In units of frequency, the response of a link to the GW signal is given as \cite{maggiore2008gravitational}:
 \begin{equation}
\Delta \nu_h=\frac{\nu_0}{2(1-\hat{k}_{GW} \cdot \hat{n})}\left(\mathbf{h}_{ij}\left(t_e -\frac{\hat{k}_{GW} \cdot \vec{r}_e}{c}\right)-\mathbf{h}_{ij}\left(t_r - \frac{\hat{k}_{GW}\cdot \vec{r}_r}{c}\right)\right)n^i n^j,
\label{formula12}
\end{equation}
where $\hat{k}_{GW}$ is the versor that describes the direction of propagation of the gravitational wave from the source to the detector, $\nu_0$ is the nominal laser frequency, $\hat{n}$ stands for the LISA link versor, $t_e$ and $t_r$ are the emission and reception time of the beam, respectively, and $\vec{r}_e$ and $\vec{r}_r$ are the satellites position vectors. All these quantities are expressed in the Solar System reference frame. \\ Let us introduce some nomenclature that will be useful in the next paragraph. The symbol $x_{j,i}$, as illustrated in Fig. \ref{LISA}, represents the time-series of the relative displacement of the TM located inside satellite $j$ and at the end of the link starting from satellite $i$. The relative displacement of the TM is considered along the direction of the link and it is relative to the local inertial frame.
Within the assumptions given before and accordingly to the center of mass (COM) reference frame $\hat{x}-\hat{y}$ in Fig. \ref{LISA}, we can compute the interferometric signals in the six LISA links that can be described in frequency
as illustrated in TABLE \ref{tabledeltx}. \\ To do the calculation, we consider the approximation that all delays are constant and equal to $\tau$. As we will see in section \ref{nneqarm}, this assumption holds well for some combinations, like the classic Michelson X, while others deviate significantly in the more realistic case of unequal armlengths. Still, it is a good working assumption to describe the single link signals, especially in the low-frequency regime.\\

\begin{table}
\begin{center}
\begin{tabular}{|c|c|}
\hline 
Link & $\Delta \nu$  \\ \hline
2 $\rightarrow$ 1 & $\frac{\nu_0(t)}{c}\left(\dot{x}_{1,2}(t) + \dot{x}_{2,1}(t-\tau)\right)-\frac{ \nu_0(t)}{4}(h_+(t)-h_+(t-\tau))+\frac{ \nu_0(t)\sqrt{3}}{4}(h_\times(t)-h_\times(t-\tau))$ \\ \hline
1 $\rightarrow$ 2 & $\frac{ \nu_0(t)}{c}\left(\dot{x}_{1,2}(t-\tau) + \dot{x}_{2,1}(t)\right)-\frac{ \nu_0(t)}{4}(h_+(t)-h_+(t-\tau))+\frac{ \nu_0(t)\sqrt{3}}{4}(h_\times(t)-h_\times(t-\tau))$  \\  \hline
3 $\rightarrow$ 1 & $\frac{ \nu_0(t)}{c}\left(\dot{x}_{1,3}(t) + \dot{x}_{3,1}(t-\tau)\right)-\frac{ \nu_0(t)}{4}(h_+(t)-h_+(t-\tau))-\frac{ \nu_0(t)\sqrt{3}}{4}(h_\times(t)-h_\times(t-\tau))$  \\    \hline
1 $\rightarrow$ 3 & $\frac{ \nu_0(t)}{c}\left(\dot{x}_{1,3}(t-\tau) + \dot{x}_{3,1}(t)\right)-\frac{ \nu_0(t)}{4}(h_+(t)-h_+(t-\tau))-\frac{ \nu_0(t)\sqrt{3}}{4}(h_\times(t)-h_\times(t-\tau))$   \\   \hline
3 $\rightarrow$ 2 & $\frac{ \nu_0(t)}{c}\left(\dot{x}_{3,2}(t-\tau) + \dot{x}_{2,3}(t)\right)+\frac{ \nu_0(t)}{2}(h_+(t)-h_+(t-\tau))$  \\ \hline
2 $\rightarrow$ 3 & $\frac{ \nu_0(t)}{c}\left(\dot{x}_{3,2}(t) + \dot{x}_{2,3}(t-\tau))\right)+\frac{ \nu_0(t)}{2}(h_+(t)-h_+(t-\tau))$  \\  \hline
\end{tabular}
\end{center}
\caption[]{\footnotesize{Table of the six single LISA links expressed in units of frequency for a GW wave propagating normal to the LISA plane. The symbol $x_{j,i}$ is the small signal displacement  of the TM located inside satellite $j$, and at the end of the link starting from satellite $i$, and $\dot{x}_{j,i}$ is its derivative. $h_+$ and $h_\times$  are the plus and cross polarization of the GW signal, and their respective coefficients follow from the projection of the GW along the LISA link. 
} }
\label{tabledeltx}
\end{table} 

These assumptions allow us to use linear algebra to find all combinations of the 210 TDI combinations which are null to GWs and can be used to measure the instrumental noise. See Supplemental Material at [URL] for the full set of combinations null to GWs. 
 
Considering the low frequencies regime, in the approximation of constant and equal delays, the expression for the GW signal for a general $TDI_k$ combination can be computed as \footnote{Note that in defining these coefficients, we fix the time at which each TDI combination is measured. A more conclusive search would allow time shifts between the different combinations, which might allow us to find further null channels. We will explore how the TDI combinations relate to each other under timeshifts in a follow-up article \cite{Olafetall_2021}, where we will demonstrate that such time shifts do not help in recovering additional information.}:
\begin{equation}
h[TDI_k](t) = \sum_{j=0}^{8}(a^{+}_{k,j}h_+(t-j \tau) +a^{\times}_{k,j}h_\times(t-j \tau)) \equiv \sum_{j=1}^{18}\alpha_{k,j}h_j(t).\label{heq}
\end{equation}
Here, $a^{+}_{k,j}$ and $a^{\times}_{k,j}$ are the coefficients of $h_+$ and $h_\times$, computed using TABLE \ref{tabledeltx} for all the 210 combinations. 
Since we have 9 coefficients for each of the $+$ and $\times$ polarizations, we count a total of 18 $\alpha_{k,j}$ coefficients for each combination.\\

We also calculate the signal due to acceleration of the TMs which has the form:
 \begin{equation}
g[TDI_k](t) = \sum_{j=0}^{8} \sum_{i\neq n=1}^{3}b_{k i n j} g_{i,n}(t-j \tau) \equiv  \sum_{j=1}^{54} \beta_{kj} g_{j}(t),
\end{equation}
where $b_{k i n j}$ are the coefficients of the six TMs acceleration terms, $g_{i,n}=\ddot x_{i,n}$, computed for each time $t-j \tau$. The equation has been computed considering that each TM has two indices: the first one stands for the satellite $i$ the TM is located in, while the second one, $n$, stands for which satellites the TM faces. Both indices $i$ and $n$ can go from 1 to 3, with the restriction that $i\neq n$. Note that $g_{i,n}\neq g_{n,i}$, as these terms describe different TMs. Similar to  Eq. (\ref{heq}), there are nine measuring time, such that we have a total of 54 coefficients  $\beta_{kj}$ for the six TMs.

We can now look for linear combinations of the combinations that form Eq. (\ref{heq}), which we call $\psi_l(t)$, that are insensitive to GWs:
\begin{equation}
\psi_l(t) =\sum_{k=1}^{N}  c _{lk} h[TDI_k](t)= 0,\label{psieq}
\end{equation}
where $N=210$ is the total number of laser noise suppressing TDI combinations. There are multiple solutions for $\psi_l(t)$, which are distinguished by the label $l$. 

We can insert Eq. (\ref{heq}) into Eq. (\ref{psieq}) and consider each $h_j$ to be independent to conclude that $\psi_l(t)$ is zero under the condition that
\begin{equation}
\sum_{k=1}^{N}c_{lk}\alpha_{k,j}=0.
 \label{eq:calpha}
\end{equation}
Thus, any row $l$ of the null space of the matrix $\alpha_{k,j}^T$ contains the coefficients $c_{lk}$ of one possible linear combination of the 210 TDIs null to GW. \\
We find that the null space matrix has 200 rows, i.e. there are 200 possible combinations that cancel the GW signal. This means that there are 200 ways of combining our 210 laser noise cancelling TDI combinations to find new combinations that give no GW signal.

Once the null space $c_{lk}$  has been found, we can calculate the TM acceleration signals $g[\psi_l](t)$ for each combinations $l$ null to GW as following:
\begin{equation}
g[\psi_l](t) =\sum_{k=1}^{N}c_{lk}\sum_{j=1}^{54}\beta_{kj}g_j(t).
\label{gpsi}
\end{equation}
It turns out that for some value of $l$ these combinations are also null to acceleration noise, that is, $\sum_{k=1}^{N}c_{lk} \beta_{kj} = 0$. Out of the 200 combinations, we find a total of 24 combinations made of linear combinations of 16-links combinations that in the equal arms configuration give no output. This means that at least in the approximation of equal arms, these combinations do not measure anything, and we discard them for the further analysis. We will investigate the underlying reason for this effect in a follow up work \cite{Olafetall_2021}. \\
We will continue our analysis using only the remaining 176 combinations, and to which we will refer in the following as the null-channels \footnote{
To exclude the possibility that it might be due to our approximation of considering equal and constant arm-lengths, we also checked their behaviour for a non-equal arm-length constellation and get the same result. Therefore we conclude that these combinations are mathematical artefacts and do not seem to have a practical purpose.}. We remark that combination number 176 of that list is the T combination that is given as the sum of the three Michelson combinations. Notice also that, out of these 176 null channels, there are 21 which are single TDI combinations that already suppress the GW signal on their own. In particular these are $C^{12}_3$, $C^{14}_3$, $C^{16}_{26}$ and $C^{16}_{27}$ plus their  satellites permutations symmetries. Of the remaining 155 null channels, 127 are made of linear combinations of only 16-link combinations, 18 are made of linear combinations of 16-link and 14-link combinations, and 7 are made of linear combinations of 16-link and 12-link combinations.\\

Furthermore, we find that the rank of the matrix  of coefficient $\delta_{ij}$ defined as:
\begin{equation}
\delta_{ij}=\sum_{k=1}^{N}c_{ik}\beta_{kj},
\label{nullg}
\end{equation}
 is 7. Thus, most of the null-channels which have a non zero acceleration signal contain redundant information. In the next section we follow  the method of the single value decomposition to decompose the matrix and see if we are able to retrieve the single TM acceleration noises. 

 \section{\label{svds1} Single value decomposition to retrieve the acceleration noise signals}
 
As we found out  that the rank of the null space is 7, we can continue our search  considering a reduced set of combinations, which will improve the computation time of our algorithm. We construct a set of 102 combinations by using each core combination and its two cyclic permutations. Considering these symmetries ensures that we include combinations starting at each of the satellites and that we use all of the inter-satellite links. This reduced set then yields 92 null channels free of GW signals. The combinations $C^{16}_4$, $C^{16}_{24}$, $C^{16}_{26}$, $C^{16}_{27}$, $C^{16}_{28}$, $C^{14}_3$, $C^{12}_3$ are null by themselves, such that including their two cyclic permutations they account for 21 out of the 92 null channels. Of the remaining 71 null channels, 59  are combinations of 16 link combinations, 6 are combinations of the 16 and 14 links combinations, and finally 6 are combinations of the 16 and 12 links combinations. \\ As we saw in the previous section, there are six TMs and nine measuring times, such that the acceleration signals enter with 54 distinct coefficients. To retrieve which of the combinations are  independent, we can do  the singular value decomposition of the $92 \times 54$ matrix as explained in Appendix \ref{svds}. The results is a combination of the acceleration coefficients $\beta_{kj}$ which we call $g[\psi_m] (t)$. These combinations of combinations are represented by:
\begin{equation}
g[\psi_m] (t)=\sum_{i=1}^{N_{\text{null}}}U_{mi}^{-1}\sum_{k=1}^{N_{TDI}}c_{ik}\sum_{j=1}^{54}\beta_{kj}g_j(t),
\end{equation}
where $N_{\text{null}}=92$ is the number of null channels, $N_{TDI}=102$ the total number of TDI combinations we considered and $c _{lk}$ are the coefficients that belong to the null-space of the matrix, while $U_{mi}$ is a $92 \times 92$ matrix. \\
We empirically found that the 54 $g_{j}(t)$ elements, after this operation, appear only in the following form:
\begin{equation}
g[\psi_m] (t)=\sum_{n=0}^8\epsilon_{mn}G_n(t).\label{gspieq}
\end{equation}
 The index $m$ goes from 1 to 7 as we have 7 independent signals, and $n$ goes from 0 to 8 as we have 8 delays at maximum. The 9 combinations $G_n(t)$ that appear are combinations of the following quantities for any value of the coefficient $n$:
 \begin{equation}
 \begin{split}
 G_n(t) = & g_{1,2}(t-n\tau)-  g_{2,1}(t-n\tau)+ g_{3,1}(t-n\tau) \\
 &-g_{1,3}(t-n\tau)+ g_{2,3}(t-n\tau) -g_{3,2}(t-n\tau). \label{rotsig}
 \end{split}
 \end{equation}

It is then useful to compare these signals $G_n(t)$ with the acceleration noise $g_{X}(t)$ that enters in the X combination. Let us first write the acceleration signal for the TDI X in units of displacement as following:
\begin{equation}
X(t)= \Delta_{X_0}(t) -  \Delta_{X_2}(t) -  \Delta_{X_4}(t)+  \Delta_{X_6}(t).
\label{deltaxdX}
\end{equation}
 Here we have defined the fundamental quantity as:
 \begin{equation}
 \begin{split}
 \Delta_{X_n}(t) =&  x_{1,3}(t-n\tau)+ x_{1,3}(t-2\tau-n\tau)+2 x_{3,1}(t-\tau-n\tau)\\ & -( x_{1,2}(t-n\tau)+ x_{1,2}(t-2\tau-n\tau)+2 x_{2,1}(t-\tau-n\tau) ),
  \end{split}\label{eq:accX}
 \end{equation}
 with $n = {0,2,4,6}$. Thus, TDI X measures a combination of the fundamental quantity given by Eq.~\ref{eq:accX}. Expanding Eq.~\ref{deltaxdX} to leading order in $\tau$ and considering that the second derivative of the displacement $x_{j,i}(t)$ is the TM acceleration $g_{j,i}(t)$, we get
 \begin{equation}
 g_{X}(t) \approx 16 \tau^2 (g_{1,3}(t)+g_{3,1}(t)-g_{1,2}(t)-g_{2,1}(t)).
\end{equation}
Such that we can say TDI X has a response proportional to
 \begin{equation*}
g_{1,3}(t)+g_{3,1}(t)-g_{1,2}(t)-g_{2,1}(t).
\end{equation*}

\begin{figure}
\centering
  \includegraphics[width=0.6\linewidth]{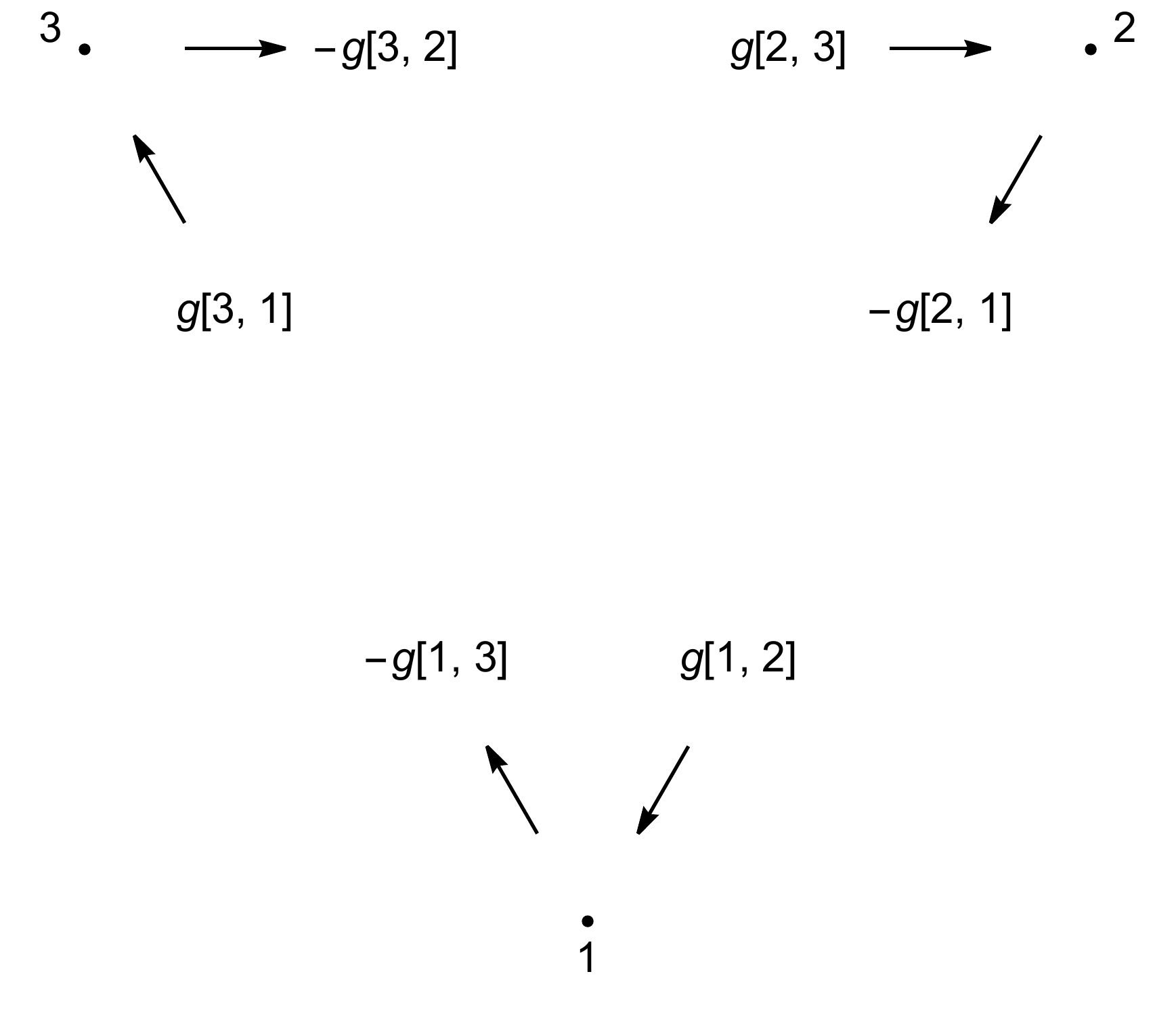}
 \caption{\footnotesize{Acceleration signal for the null combinations projected along the LISA arms. All null combinations in the case of a simplified constant equal arm-lengths configurations  behave as an ideal Sagnac interferometer.
}}
\label{rotation}
\end{figure}

\begin{figure}
\centering
  \includegraphics[width=0.5\linewidth]{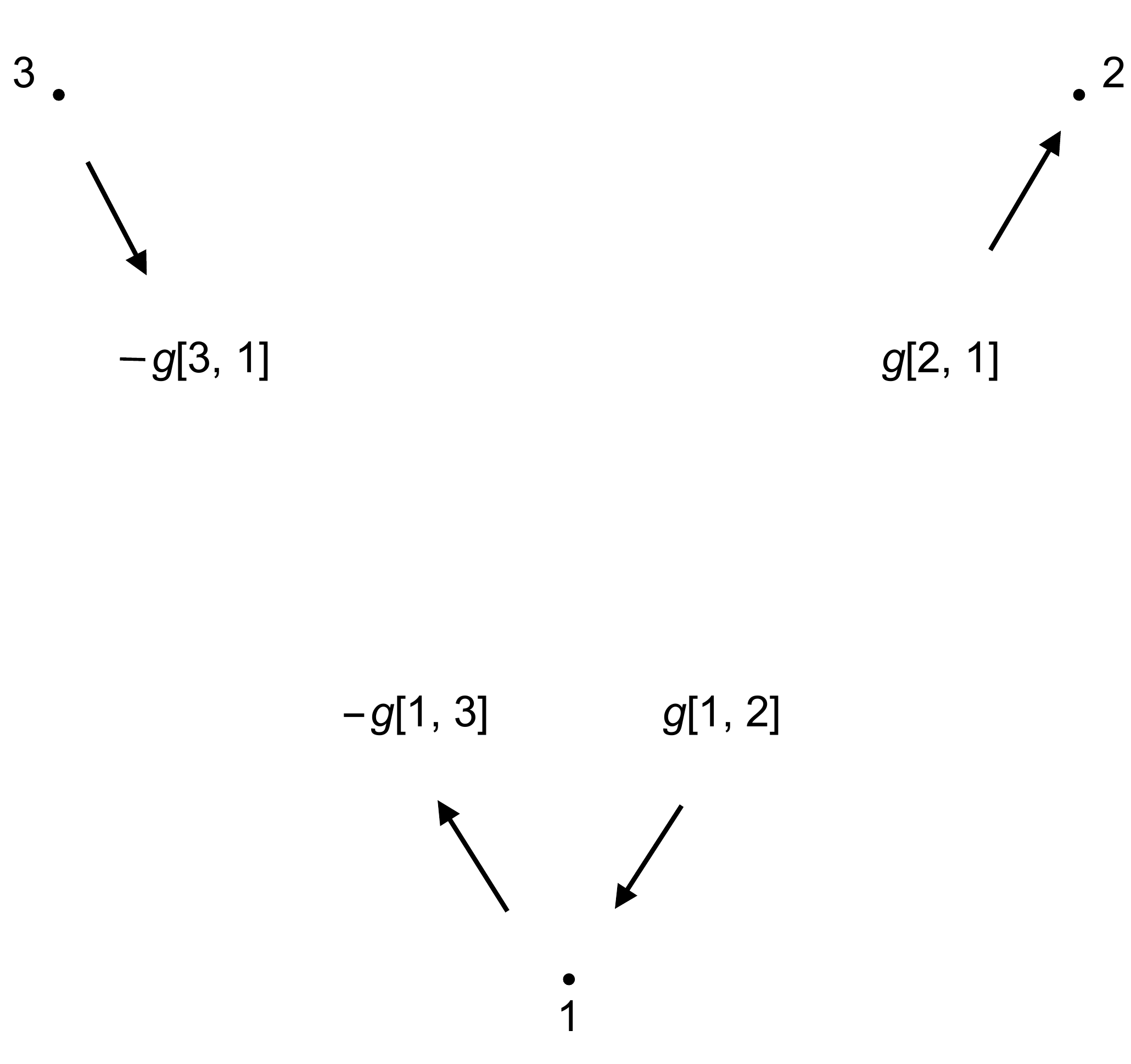}
 \caption{\footnotesize{Acceleration signal for the TDI X projected along the LISA arms. The TDI X in the case of a simplified constant equal arm-lengths configurations is sensitive to the a-mode of a triangle.
}}
\label{modeX}
\end{figure}

The symmetry of the Eq. (\ref{rotsig}) it's worth investigating.  If we simplify the problem and reduce the six LISA TMs to three TMs, that coincide with the vertex of a triangle, one can decompose the acceleration $g_{j,i}(t)$  into linear combinations of the modes of an equilateral spring-mass triangle, and readily calculate that only rotation has a non zero projection onto $G_n(t) $ as visible in Fig. \ref{rotation}. Conversely, the Michelson combination X shows to be sensitive to a different mode of the triangle that is displayed in Fig. (\ref{modeX}). Notice that,  this is the \textit{a-mode} and it is orthonormal to the rotation mode.
\noindent In other words, considering a perfect drag-free attitude control system and that each spacecraft is rigidity connected to the TMs, if we rotate the constellation around the LISA COM, the type of signal we see, in first approximation, can be considered as representing the rigid rotation of the constellation. This is the only thing we can measure in this set up, which implies that if we loose the laser link in one arm we would also loose our ability to construct a null channel.

The corresponding coefficients' matrix $\epsilon_{mn}$ ($ 7 \times 9$)  for each time value ($n$ goes from 0 to $9$) of $G_n(t)$ is reported in Table \ref{epsilonmn1}.
 \begin{table}
\centering
\Large
 \begin{tabular}{|c|c|c|c|c|c|c|c|c|c|}
 \hline
 & $G_0$ & $G_1$ & $G_2$ & $G_3$ & $G_4$ &  $G_5$ & $G_6$ & $G_7$ & $G_8$ \\ \hline
 $\psi _1$ & $\frac{137}{12}$ & $-\frac{464}{25}$ & $-\frac{142}{23}$ & $\frac{314}{19}$ &$ \frac{140}{33}$ & $-\frac{285}{32}$ & $\frac{317}{222}$ & $-\frac{53}{508}$ & $\frac{110}{833}$ \\
$\psi _2$ & $-\frac{105}{37}$ & $\frac{96}{25}$ & $-\frac{34}{9}$ & $\frac{157}{17}$ & -$\frac{356}{37}$& $\frac{155}{43}$ & $\frac{39}{100}$ & $-\frac{119}{149}$ & $-\frac{5}{144}$ \\
$\psi _3$ & $\frac{74}{75}$ & $-\frac{49}{12}$ & $\frac{14}{3} $ & $-\frac{7}{9}$ & -3 & $\frac{11}{3} $ & $ -\frac{3}{10}$ &$ -\frac{8}{5}$ &$ \frac{94}{203}$ \\
 $\psi _4$ & $-\frac{89}{39}$ &$ \frac{32}{153}$ & $\frac{298}{77}$ & $\frac{75}{61}$ & -$\frac{124}{111}$ & $-\frac{134}{33}$ & $\frac{69}{44} $ & $\frac{79}{84}$ & $-\frac{109}{305}$ \\
 $\psi _5$ & $-\frac{199}{140}$ & $-\frac{55}{227} $ & $\frac{94}{147}$ & $\frac{127}{109} $&
 $  \frac{49}{37}$ & $\frac{48}{71}$ & $-\frac{41}{19}$ &$ \frac{59}{149} $ & $-\frac{47}{124}$ \\
$ \psi _6$ & $-\frac{55}{107} $ &$ \frac{21}{86} $ & $\frac{3}{31}$ & $\frac{125}{368} $& $\frac{135}{212} $&
  $ -\frac{73}{328}$ & $\frac{116}{567}$ & $-\frac{223}{103}$ & $\frac{109}{79}$ \\
$ \psi _7$ &$ -\frac{47}{124} $&$ -\frac{57}{370} $& $-\frac{16}{279}$ &$ \frac{34}{299}$ &
   $\frac{59}{156}$ & $\frac{82}{217}$ & $\frac{287}{383}$ & $-\frac{7}{27} $&$ -\frac{227}{295}$ \\ \hline
\end{tabular}
\caption{Table of the $\epsilon_{mn}$ coefficients for the 7 independent combinations in case of LISA with constant and equal arm-length.} \label{epsilonmn1}
\end{table}

However, it is possible to further simplify the problem by taking the difference of $G_0$, $G_1$, $G_2$, $G_3$, $G_4$, $G_5$, $G_6$, $G_7$ and $G_8$ i.e. the differences of the acceleration signals. 
As visible in Table \ref{table:epsilonmn}, we are then able to further reduce the matrix $\epsilon_{mn}$ from a $7 \times 9$ matrix to a $7 \times 7$ matrix, using the backward finite differences of $G_n(t)$  \footnote{In detail, we have \begin{center}
$\begin{array}{l}
\Delta G_0=G_0 \\
\Delta G_1=G_0-G_1 \\
\Delta G_2=G_0-2 G_1+G_2 \\
\Delta G_3=G_0-3 G_1+3 G_2-G_3 \\
\Delta G_4=G_0-4 G_1+6 G_2-4 G_3+G_4\\
\Delta G_5=G_0-5 G_1+10 G_2-10 G_3+5 G_4-G_5 \\
\Delta G_6=G_0-6 G_1+15 G_2-20 G_3+15 G_4-6 G_5+G_6 \\
\Delta G_7=G_0-7 G_1+21 G_2-35 G_3+35 G_4-21 G_5+7 G_6-G_7   \\
\Delta G_8=G_0-8 G_1+28 G_2-56 G_3+70 G_4-56 G_5+28 G_6-8 G_7+G_8 \\
\end{array}.$
\end{center}},
\begin{equation}
 \Delta G_n = (-1)^k \sum_{k=0}^{8}\binom{n}{k}G_k.\label{eq:finitediff}
 \end{equation}
  \begin{table}
\centering
\Large
 \begin{tabular}{|c|c|c|c|c|c|c|c|c|c|}
 \hline
 & $\Delta G_0$ & $\Delta G_1$ & $\Delta G_2$ & $\Delta G_3$ & $\Delta G_4$ &  $\Delta G_5$ & $\Delta G_6$ & $ \Delta G_7$ & $\Delta G_8$ \\ \hline
$ \psi _1$ & 0 & 0 & $\frac{8}{3} $& $\frac{70}{3}$ & $-\frac{40}{3}$ & $-\frac{34}{7}$ & $\frac{22}{5}$ &
  $ -\frac{20}{21}$ & $\frac{7}{53}$ \\
 $\psi _2$ & 0 & 0 &$ -\frac{19}{2}$ & $\frac{46}{3}$ & $-\frac{113}{7}$ & $\frac{51}{4}$ & $-\frac{37}{6}$
   & $\frac{14}{13}$ & $-\frac{5}{144}$ \\
$ \psi _3$ & 0 & 0 & -4 & 12 & $-\frac{25}{2} $ & $\frac{17}{3}$ & $\frac{3}{2}$ & -$\frac{19}{9}$ &
   $\frac{19}{41}$ \\
$ \psi _4$ & 0 & 0 & $-\frac{13}{2}$ & $-\frac{1}{2}$ & 10 & $-\frac{56}{11}$ & $-\frac{13}{7}$ &
   $\frac{23}{12} $&$ -\frac{5}{14}$ \\
$ \psi _5$ & 0 & 0 & $-\frac{79}{5} $ & $\frac{112}{3}$ & $-\frac{121}{3}$ & $\frac{126}{5}$ & -10 &
 $  \frac{29}{11}$ & $-\frac{11}{29} $\\
 $\psi _6$ & 0 & 0 & -1 & $-\frac{25}{4}$ & $\frac{47}{2}$ & $-\frac{131}{4} $ & $\frac{71}{3} $&
   $-\frac{71}{8}$ & $\frac{29}{21}$ \\
 $\psi _7$ & 0 & 0 & $-\frac{47}{5}$ & $\frac{127}{4}$ & $-\frac{99}{2}$ & $\frac{131}{3} $&
  $ -\frac{113}{5}$ & $\frac{77}{12}$ & $-\frac{10}{13} $ \\ \hline
\end{tabular}
\caption{Table of the $\epsilon_{mn}$ coefficients for the 7 independent combinations in case of LISA with constant and equal arm-length via the new difference variable $\Delta G_n$. Note that the relation can be inverted to give the $\Delta G_k$ for $k \geq 2$} \label{table:epsilonmn}
\end{table} 
 
Thus, we can reduce the set of 9 signals from $G_0$ to $G_8$ by using these new difference variables and obtain an invertible matrix made out of the $\Delta G_n$.
 Notice that switching to $\Delta G_n$ causes the term $\Delta G_0$ and $\Delta G_1$ to disappear.  So, in this way, if we use the analogy with a backward finite differences of $G_n(t)$ the minimum quantity we can measure is a second derivative of the $\Delta G$ and the relation can be inverted to give the $\Delta G_k$ for $k \geq 2$.\\


However, since the rank of the previously obtained null space is 7, we do not really need to start our analysis using such a large set of TDI combinations, and we can try to reduce the set of 102 combinations we are using even further. Therefore, instead of building the null-channels by linearly combining the three copies of the original core obtained by applying the two cyclic spacecraft indices permutations to the 34 core combinations (listed in Tables \ref{T16l}, \ref{T14l} and \ref{T12l}), we applied the same criterion that is currently used to build the combination T for all core TDI combinations. That is: we produce a new set of combinations $\Tilde{C}_k$ by averaging the signals over the combinations obtained by applying the two cyclic permutations of the satellites to the 34 core combinations. 
Note that an advantage of this approach is that we directly obtain a new set of 34 combinations which are null to GWs  in the case of a perfect equilateral LISA, so we do not have to look for the null space \footnote{It might also help for an actual implementation of these combinations, since each of these null-channels is built out of combinations which belong to the same topology. Indeed, for the combinations T, we had just to sum the X,Y,Z without applying any additional delays. This means we can construct them without considering relative timeshifts between combinations of different topology.}.\\
Furthermore, we consider here only 31 combinations out of the 34 since we exclude three core combinations that we found to not bring any informations when considering equal arms that are $C_4^{16}, C_{24}^{16}$ and $C_{28}^{16}$. Indeed as we will show in detail in a follow up work \cite{Olafetall_2021}, these combinations in case of constant and equal arm-length give no output, while they strongly suppress both noise and GW signals in the realistic case of unequal, time-varying arms.\\  

It turns out that there are many possible independent combinations we can form. 
However, from empirical studies we found that the ones that have the minimum number of links involved yield simpler expressions for estimating the acceleration noise, while containing the same information as any other set. These TDIs channels are $\tilde{\mathcal{C}}^{16}_{1}$, $ \tilde{\mathcal{C}}^{16}_{2}$, $ \tilde{\mathcal{C}}^{14}_{1}$, $ \tilde{\mathcal{C}}^{14}_{3}$, $ \tilde{\mathcal{C}}^{12}_{1}$, $ \tilde{\mathcal{C}}^{12}_{2}$, $ \tilde{\mathcal{C}}^{12}_{3}$. In Table \ref{epsilonmn} we compute the acceleration signals for this set.\\
  \begin{table}
\centering
 \begin{tabular}{|c|c|c|c|c|c|c|c|c|c|}
 \hline
$\begin{array}{c|c|c|c|c|c|c|c|c|c}
 \text{} & \text{$\Delta $G}_0 & \text{$\Delta $G}_1 & \text{$\Delta $G}_2 &
   \text{$\Delta $G}_3 & \text{$\Delta $G}_4 & \text{$\Delta $G}_5 & \text{$\Delta $G}_6
   & \text{$\Delta $G}_7 & \text{$\Delta $G}_8 \\ \hline
 \tilde{\mathcal{C}}^{16}_{1} & 0 & 0 & 0 & 0 & -\frac{8}{3} & \frac{16}{3} & -\frac{14}{3} & 2
   & -\frac{1}{3} \\
 \tilde{\mathcal{C}}^{16}_{2} & 0 & 0 & -2 & 6 & -\frac{32}{3} & \frac{34}{3} & -7 & \frac{7}{3}
   & -\frac{1}{3} \\
 \tilde{\mathcal{C}}^{14}_{1} & 0 & 0 & -2 & 3 & -\frac{5}{3} & \frac{1}{3} & 0 & 0 & 0 \\
 \tilde{\mathcal{C}}^{14}_{3}  & 0 & 0 & -2 & 1 & 0 & 0 & 0 & 0 & 0 \\
 \tilde{\mathcal{C}}^{12}_{1} & 0 & 0 & -3 & 6 & -5 & 2 & -\frac{1}{3} & 0 & 0 \\
 \tilde{\mathcal{C}}^{12}_{2}& 0 & 0 & -1 & 1 & -\frac{1}{3} & 0 & 0 & 0 & 0 \\
 \tilde{\mathcal{C}}^{12}_{3} & 0 & 0 & -1 & 0 & 0 & 0 & 0 & 0 & 0 \\ \hline
\end{array}$
\end{tabular}
\caption{Matrix of coefficients of the acceleration signals for the set $ \tilde{\mathcal{C}}^{16}_{1}$, $ \tilde{\mathcal{C}}^{16}_{2}$, $ \tilde{\mathcal{C}}^{14}_{1}$, $ \tilde{\mathcal{C}}^{14}_{3}$, $ \tilde{\mathcal{C}}^{12}_{1}$, $ \tilde{\mathcal{C}}^{12}_{2}$, $ \tilde{\mathcal{C}}^{12}_{3}$. The matrix can be inverted to give the $\Delta G_k$ for $k \geq 2$.} \label{epsilonmn2}
\end{table} 
The matrix of coefficients from Table \ref{epsilonmn2} can be then inverted to give the $\Delta G_k$ for $k \geq 2$ as:\\ 
\begin{subequations}
\begin{align}
\text{$\Delta $G}_2&=-\tilde{\mathcal{C}}_3^{12}\\
\text{$\Delta
   $G}_3&=\tilde{\mathcal{C}}_3^{14}-2 \tilde{\mathcal{C}}_3^{12}\\
   \text{$\Delta $G}_4&=-3
   \left(\tilde{\mathcal{C}}_2^{12}+\tilde{\mathcal{C}}_3^{12}-\tilde{\mathcal{C}}_3^{14}
   \right)\\
   \text{$\Delta $G}_5&=3 \left(\tilde{\mathcal{C}}_1^{14}-5
   \tilde{\mathcal{C}}_2^{12}-\tilde{\mathcal{C}}_3^{12}+2
   \tilde{\mathcal{C}}_3^{14}\right)\\
   \text{$\Delta $G}_6&=-3
   \left(\tilde{\mathcal{C}}_1^{12}-6 \tilde{\mathcal{C}}_1^{14}+15
   \tilde{\mathcal{C}}_2^{12}-3 \tilde{\mathcal{C}}_3^{14}\right)\\
   \text{$\Delta $G}_7&=-3
   \left(7 \tilde{\mathcal{C}}_1^{12}-24
   \tilde{\mathcal{C}}_1^{14}+\tilde{\mathcal{C}}_1^{16}+39
   \tilde{\mathcal{C}}_2^{12}-\tilde{\mathcal{C}}_2^{16}-4 \tilde{\mathcal{C}}_3^{12}-3
   \tilde{\mathcal{C}}_3^{14}\right)\\
   \text{$\Delta $G}_8&=-3 \left(28
   \tilde{\mathcal{C}}_1^{12}-76 \tilde{\mathcal{C}}_1^{14}+7
   \tilde{\mathcal{C}}_1^{16}+96 \tilde{\mathcal{C}}_2^{12}-6
   \tilde{\mathcal{C}}_2^{16}-16 \tilde{\mathcal{C}}_3^{12}\right).
\end{align}\\
\end{subequations}
\noindent Note that $\Delta G_2$ is directly given as $-\tilde{\mathcal{C}}_3^{12}$, which turns out to be directly proportional to $\mathcal{C}_3^{12}$ (i.e. the fully symmetric Sagnac combination $\zeta$), at least within our approximation. See \cite{Muratore:2021phd} for more information.  
Moreover, $\Delta G_3$ up to $\Delta G_6$  can be estimated using just 12-link and 14-link combinations.\\
According to Eq. \ref{eq:finitediff},  $\Delta G_k$ with $k>0$ are finite differences of the same quantity, and we have shown here that the first $\Delta G_k$ we can measure is $\Delta G_2$. Finite differences are just linear operators, like time derivatives, and they would not affect the signal to noise ratio or the ratio between the responses to different signals such as GW and glitches. Therefore, since all the $\Delta G_k$ with $k>2$, are higher order finite differences of the same signal $\zeta$, every null channel you can build using up to 16 links long TDI, gives at most the same information as $\zeta$.\\ 
In the next paragraph to validate the beforehand conclusion we translate this to a more realistic LISA model and assess the sensitivity for the null-channels $\zeta$ and T.
 
 \section{\label{nneqarm} Computation of the effective h noise for a perfect LISA triangle and a non perfectly equilateral triangle}
 
\begin{figure}
\begin{center}
\includegraphics[width=0.8\linewidth]{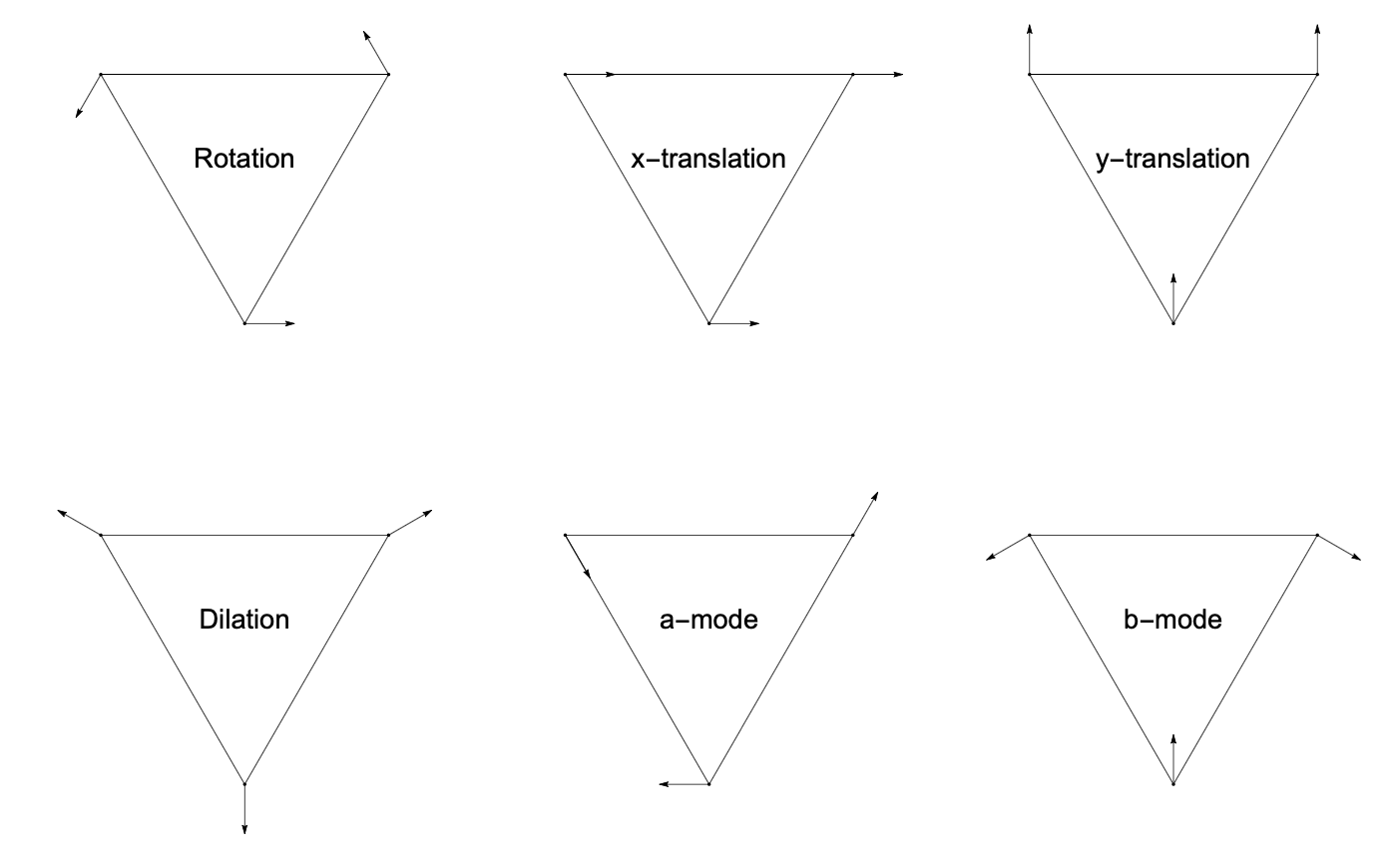}
\caption{Normal modes of an equilateral spring-mass triangle. Figure adapted form ref. \cite{normalmodes}.}
\label{triangle}
\end{center}
\end{figure}

 Let us now relax the hypothesis of a perfectly equilateral LISA and compute the TM acceleration noise  considering constant but not equal arm-lengths. \\
Since LISA is a constellation of satellites that form ideally an equilateral triangle, we decided to approach the problem of the inequality of the arm-lengths by assuming that LISA undergoes small amplitude static distortions along any of the triangle normal modes that are illustrated in Fig. \ref{triangle}.\\
We can then study the impact of the distortion modes following \cite{normalmodes}, and calculate which one causes an effective distortion with respect to the nominal LISA. 
To linear terms in the distortion, only the \textit{a-mode}, \textit{b-mode} and \textit{d-mode} (dilation) cause a change in the length of the arms. The dilatation mode shows to distort LISA equally in all the directions, such that it does not modify the assumption of equal arms. Whereas, the \textit{a-mode} and \textit{b-mode}, cause a differential change between the arms, while leaving the overall perimeter of the triangle at its nominal value.
 
We therefore just consider the \textit{a-mode} ($\delta_a$)  and \textit{b-mode} ($\delta_b$) and assume that all other modes have zero amplitude.  Then, we can express the three LISA arms, to linear terms in the distortion, as:
\begin{equation}
\begin{split}
L_{12}&= c \tau \left(1 + \frac{1}{2}\left(\sqrt{3}\delta_a - \delta_b\right)\right)\\
L_{23}& = c \tau \left(1 + \delta_b\right)\\
L_{31} &= c \tau \left(1 - \frac{1}{2}\left(\sqrt{3}\delta_a + \delta_b\right)\right).
\end{split}\label{Lterm}
\end{equation}
We define the root-mean square (RMS) deviation $\delta_r $ of the arm-length as:
\begin{equation}
\delta_r = \sqrt{\frac{\delta_a^2 + \delta_b^2}{2}}.
\end{equation}
We then consider the leading order terms in the distortion for LISA with unequal but constant arm-length, to calculate in the frequency domain the response to a GW coming orthogonal  with respect to LISA plane. We can do the same with the acceleration noise and expand both the acceleration signal and the GW signal as a function of frequency and take only the leading term at low frequencies. In this process, we assume that all $g_{i,j}(\omega)$ have the same power spectral density (PSD) $S_g(\omega)$, and calculate the effective h noise using the transfer function for $h_{rms}=\sqrt{h_+^2+h_\times^2}$,
 as we will explain in the following. We refer to effective h noise as the level of the TM acceleration noise we would be able to detect in the presence of a GW background \footnote{What we call here effective h noise is also often referred to as strain sensitivity, given by renormalizing the instrument noise PSD by the gravitational wave transfer function of the instrument, see e.g. \cite{Larson_2000}.}.\\
 
  We can compute the effective h noise by considering that the output of a TDI$_j$, given a stochastic GW background, is a stochastic process with a PSD of the type
\begin{equation}
S_{{j}_h} = |H_j(\omega)|^2S_h(\omega)\label{sh},
\end{equation}
where $S_h(\omega)$ is the PSD of the GW background and $|H_j(\omega)|$ is the absolute squared value transfer function.

We can do the same consideration for the acceleration noise and write:
\begin{equation}
S_{{j}_g} = |G_j(\omega)|^2S_g(\omega)\label{sg}.
\end{equation}
We can then compute the effective h noise as:
\begin{equation}
S_{{j}_{g/h}} = \frac{S_{{j}_g}}{ |H_j(\omega)|^2},
\end{equation}
i.e., by dividing the PSD of the acceleration noise by the transfer function of the stochastic GW background for different TDI combinations. \\ 

We analyse here two cases. In the first we consider a perfect LISA triangle and 1000 GW sources coming from multiple directions, in the second a non perfectly equilateral triangle and the GW sources coming orthogonal with respect to the LISA detector.\\
Thus, for the first case the transfer function reads
\begin{equation}
S_{{j}_h} =\sum_{i}^{N_s} |H_j^i(\omega)|^2S_{h_i}(\omega), 
\end{equation}
where $N_s$ is the number of GW sources. Here, each $S_{h_i}(\omega)$ is the PSD of the RMS of the $h_\times$ and $h_+$ polarizations of the GW. In practice, we consider both $h_\times$ and $h_+$ to be white noise processes with an amplitude spectral density (ASD) of $\frac{1}{\sqrt{Hz}}$ to compute the transfer functions $|H_j^i(\omega)|^2$.
%
In TABLE \ref{equalZT} we report this effective h noise alongside the ASD of the acceleration noise and the RMS response to a wave for an ideal LISA where we averaged over the celestial sphere for T, $\zeta$ and TDI X.\\

\begin{table}
\centering
\begin{tabular}{|c|c|c|c|c|}
\hline
 & RMS response to GW & ASD of acceleration noise & Effective h noise \\
      & for ideal LISA & noise for ideal LISA & for ideal LISA \\ \hline
       X & $ \simeq 8.87\sqrt{2} \tau^3  \omega ^3$  & $ \frac{32  \tau^2 \omega \sqrt{S_g} }{c}$ & $ \simeq  \frac{2.5 \sqrt{S_g}}{c \tau \omega^2} $\\ \hline
 T & $ \simeq  0.146 \sqrt{2} \tau^6  \omega^6$  & $\frac{8 \sqrt{2}
    \tau^4 \omega^3  \sqrt{S_g}}{c\sqrt{3}}$ & $ \simeq  \frac{32 \sqrt{S_g}}{c \tau^2 \omega ^3} $\\ \hline
   $\zeta $ &$ \simeq  0.0547\sqrt{2} \tau^4 \omega^4  $ & $\frac{\sqrt{6} \tau^2 \omega  \sqrt{S_g}}{c} $ &$  \simeq  \frac{32 \sqrt{S_g}}{c \tau^2 \omega ^3}$ \\ \hline
\end{tabular}
\caption{Effective h noise for equal-constant LISA armlength in case of averaging the square of the response for 1000 GW sources over the celestial sphere for the null combinations $\zeta$ and T vs. Michelson X.}\label{equalZT}
\end{table}
In the second case, we consider only a single stochastic source orthogonal to the LISA plane, and take the inequality of the armlengths into account. We again compute the response to a single stochastic source, to the TM acceleration noise and the effective h noise, for T, $\zeta$ and TDI X. The result is summarized in TABLE \ref{slantlisapar}. \\
\begin{table}
\centering
\begin{tabular}{|c|c|c|c|c|}
\hline
 & Response to a GW coming orthogonal& ASD of acceleration  & Effective h noise \\
     &to the plane for slant LISA & noise for slant LISA  & for slant LISA \\ \hline
      X & $8 \sqrt{6} \tau^3 \omega ^3$ &$ \frac{32  \tau^2 \omega \sqrt{S_g} }{c}$ &  $\frac{2\sqrt{2}  \sqrt{S_g}  }{\sqrt{3} c \tau \omega^2 } $\\ \hline
 T & $12 \sqrt{3} \tau^3 \omega ^3 \delta _r$ &$ \frac{8 \tau^2 \omega  \sqrt{S_g} \sqrt{18 \delta _r^2+\frac{2 \tau^4 \omega ^4}{3}}}{c}$ &  $\frac{2 \sqrt{S_g} \sqrt{\frac{2 \tau^4 \omega ^4}{3}+ 18\delta _r^2 }}{3  \sqrt{3}  c \tau \omega ^2 \delta _r} $\\ \hline
   $\zeta $ &$ \frac{3}{2} \sqrt{3} \tau^3 \omega ^3 \delta _r$ & $\frac{\sqrt{6} \tau^2 \omega  \sqrt{S_g}}{c} $&$ \frac{2 \sqrt{2} \sqrt{S_g}}{3 c \tau \omega ^2 \delta _r}$ \\ \hline
\end{tabular}
\caption{Effective h noise for non equal-constant LISA arm-length in the case of a GW coming orthogonal to the LISA plane, for the null-combinations $\zeta$ and T vs. the Michelson X. Notice that the acceleration noise terms for $\zeta$ and X are not affected by the inequality of the arms and are equal to those given in Table \ref{equalZT}.}\label{slantlisapar}
\end{table}
Considering a realistic value of $\delta_r \approx 0.006$ \footnote{We obtained the estimation of the $\delta_r$ inverting the \ref{Lterm} for $L_{12},L_{23}$ and $L_{32}$, and using the ESA orbits used in  LISANode, cf. \cite{Bayle:2019phd}.}, the effective h noise of T is much smaller than that of $\zeta$. Thus, at least at low frequencies, $\zeta$ is a better noise monitor for the case of an orthogonal GW signal and unequal arm-length constellation, since it suppresses GW signals more effectively than T.\\

\begin{figure}
\begin{center}
\includegraphics[width=0.9\textwidth]{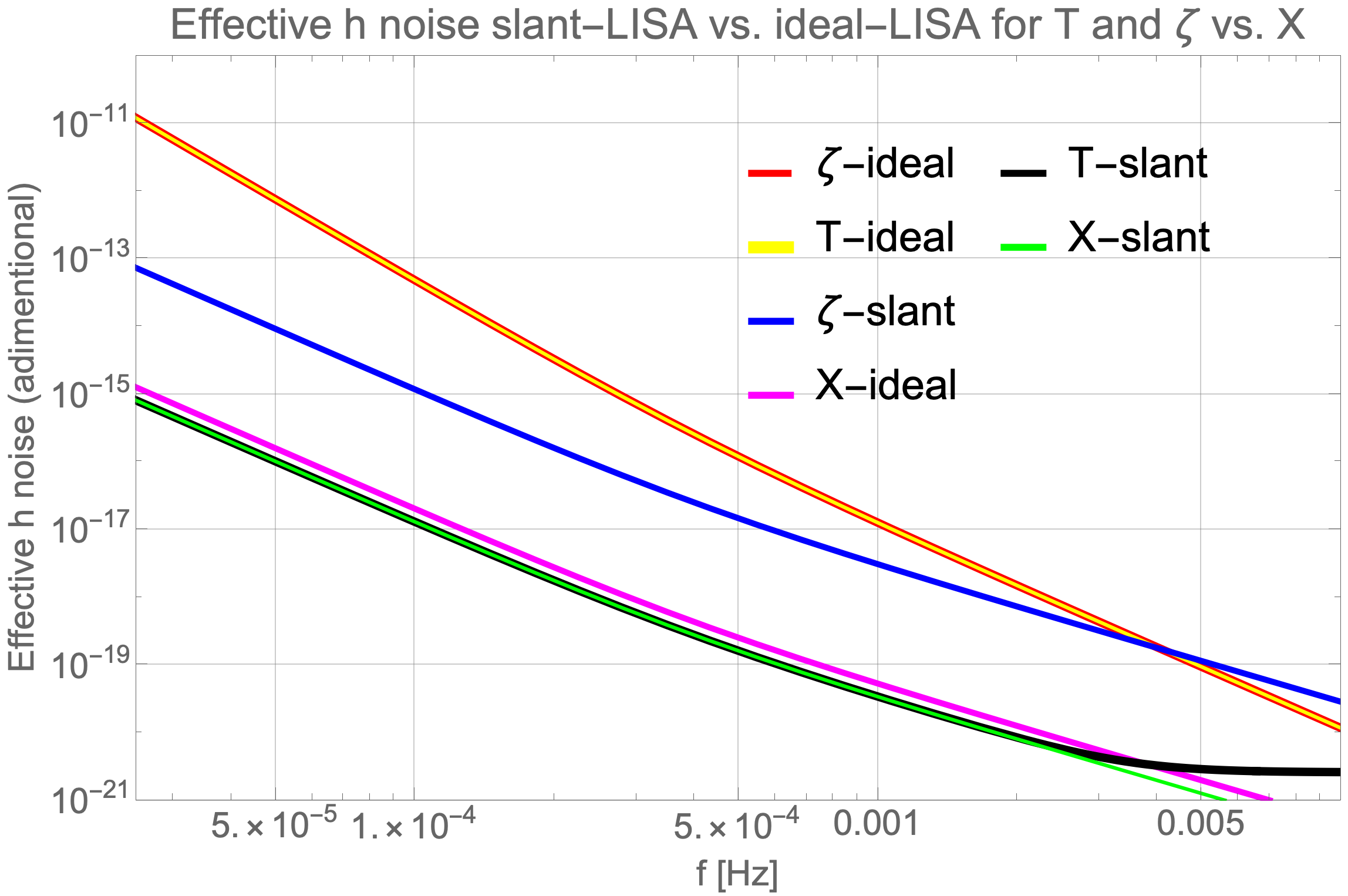}
\end{center}
\caption{\footnotesize{The effective h noise for $\zeta$ and T for both slant and ideal LISA. The graph is made considering a stochastic GW-type signal with RMS of $\sqrt{2}$ and acceleration noise with the six test masses having the same ASD equal to $S^{1/2}_g=2.4 \times 10^{-15}\sqrt{1 +\left(\frac{0.4 \times 10^{-3}}{f}\right)^2}$.}}\label{tableresults}
\end{figure}

We plot the in Fig. \ref{tableresults} the effective h noise for T, $\zeta$ and TDI X, as given in TABLE \ref{equalZT} and TABLE \ref{slantlisapar}. 

We see that
in case of a very simplistic GW background and ideal LISA (in the case of equal arms), $\zeta$ and T show to have the same sensitivity while X is much more sensitive to GW. Thus, we can state that under these assumptions both T and $\zeta$ see the same signals and are better noise monitor than X. Conversely, considering the case of LISA with unequal arm-length and with a GW source orthogonal to the plane, for frequencies lower than around \SI{E-2}{\hertz}, $\zeta$ is less sensitive to GW than T, which means that we will be able to observe the instrumental noise in the presence of larger GW signals. Indeed, at least at low frequencies, T shows to have the same sensitivity as TDI X, which is well known to be sensitive to GW signals. This is different for higher frequencies where T flattens whereas $\zeta$ shows to decrease following a constant slope. Thus, T might be a better noise monitor than $\zeta$ in this frequency region.  However, note that it is more difficult to estimate the exact combination of the acceleration noise of the six test-masses measured by T compared to the very simple rotational signal measured by $\zeta$, especially in the case of unequal arms.

\begin{figure}
\centering
  \includegraphics[width=\linewidth]{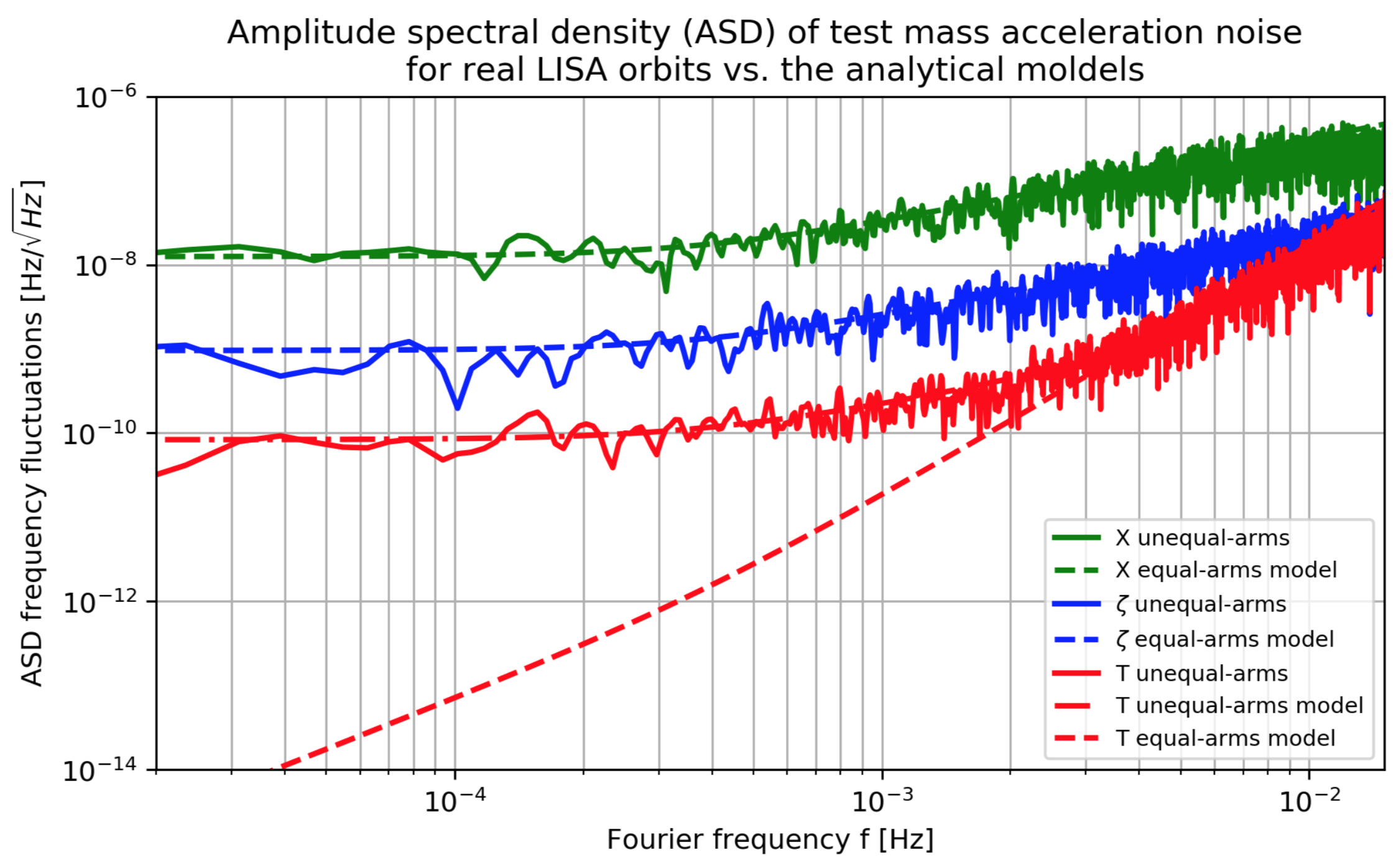}
  \caption{\footnotesize{Amplitude spectral density of the test-mass acceleration noise for the null-channel T  and the Fully-Symmetric-Sagnac $\zeta$ vs. TDI X compared with the respective analytical model. The simulations are performed considering ESA orbits }}
  \label{fig:null-combTM}
\end{figure}

For completeness, since with this paper we want to suggest which are the TDI channels to use for estimating the instrumental noise, in Fig. \ref{fig:null-combTM} we report a comparison between simulations using realistic LISA orbits, and the analytical models with ideal perfect equilateral LISA constellation, for the residual TM acceleration noise for T, $\zeta$ and X. The instrument simulation is performed using LISANode \cite{Bayle:2019phd}, while the TDI combinations are computed using the python tool pyTDI \footnote{The original codebase for pyTDI was developed by M. Staab and J.-B. Bayle, and it is now being developed by multiple contributors inside the LISA Consortium.}. The power spectral density has been estimated using the Welch method, using the \textit{Nutall4} window function.\\
 We notice that for T, the approximated analytical model that considers equal constant arm-lengths does not accurately describe the TM acceleration noise at low frequencies. This is expected, and our model accounting for inequalities in the arms given in TABLE \ref{slantlisapar} explains the observed noise level well. I.e., T no longer measures the simple signal proportional to $\Delta G_4$ (at low frequencies) when considering more realistic orbits.

\section{\label{Discussion}Discussion and conclusion}

If we go up to 16-link combinations and if we do not value the additional information that come from doing a time derivative, or a time finite difference, all information regarding the acceleration noise, measurable using null combinations, is already contained in the fully symmetric Sagnac combination $\zeta$. Thus we only need one combination to measure the properties of this noise and to characterise the instrument. \\ With the combination $\zeta$ we measure a second derivative of a particular linear combination of the test masses that resembles the rigid rotation of the constellation. Other TDI combinations insensitive to a GW signal measure combinations of higher order derivatives of the same signal.\\ 
Indeed, T in the idealised condition of a perfect triangle measures the fourth derivative.  
We have shown that in the case of a stationary, isotropic and very simplified GW stochastic background, and LISA with equal and constant arms, $\zeta$ suppresses GW equally well as the combination T. \\Regardless, T behaves differently from $\zeta$ at least from the theoretical calculation presented here, in the case of a constellation with unequal but constant arms, and a simplified GW stochastic signal coming orthogonal to the LISA plane. Indeed there are other terms in the acceleration noise that contribute and  overcome the rotational signal. In addition, we showed that $\zeta$ is significantly less sensitive to GW than T at low frequencies, when considering unequal arm-length, making it a better candidate to measure the instrumental noise.\\

Ideally, we would like to monitor the instrumental noise independently from the GW background. Unfortunately, the fact that all null combinations only measure one type of acceleration signal which is, to zeroth order, orthogonal to the one that affects TDI X (which is the combination advertised in data analysis for detecting GW signal) severely limits the possible noise correlation analyses between the X channel and the null channels.

In conclusion, we found that these null-channels are useful as LISA noise monitors. They show to be sensitive to a signal which resembles the rigid rotation of the constellation. Note that all of these null channels need all the three LISA arms to be constructed, such that in case we loose one arm of the constellation we also loose our ability to construct null channels. We are then left with combinations, such as the Michelson interferometer, which are strongly sensitive to GW signals but are not good noise monitors. We plan to translate the results presented here into a more realistic LISA noise scenario and to suggest the proper noise measurement approach for the LISA mission.

\section{Acknowlegement}
The authors would like to thank all the LISA Trento group for the useful discussion and important input on writing this paper. M.M, S.V and D.V thank also the Agenzia Spaziale Italiana and the Laboratorio Nazionale di Fisica Nucleare for supporting this work. M.M and O.H would like to thank Jean-Baptiste Bayle for useful discussions and support in verifying some of our analytical results numerically. O.H. gratefully acknowledges support by the Deutsches Zentrum für Luft- und Raumfahrt (DLR, German Space Agency) with funding from the Federal Ministry for Economic Affairs and Energy based on a resolution of the German Bundestag (Project Ref.~No.~50OQ1601 and 50OQ1801)

\newpage
\appendix

\section{\label{fullTDI} Computing the full set of TDI laser noise cancelling combinations}
The 34 combinations reported in Tables \ref{T16l},\ref{T14l},\ref{T12l} are modulo six symmetries, including the identity one. This means that we have excluded all the combinations that can be obtained from the core TDI combinations by one of these operations: 
\begin{itemize}
\item{Two cyclic permutations of the satellites:  permuting the spacecraft indices from $1\mapsto 2\mapsto 3 \mapsto 1$ or from $1\mapsto 3\mapsto 2 \mapsto 1$.}
\item{Mirror symmetry around spacecraft $i$:  interchanging the role of the other two spacecraft as $j\leftrightarrow k$, where $i,j,k$ take the values 1,2,3, and $i\neq j\neq k$.}
\item{Time reversal symmetry: switching the time direction of the sequence, i.e. inverting the direction of the arrows, so that the links switch orientation.} 
\end{itemize}

Therefore, by adding these five satellite permutation symmetries to the 34 core combinations we get a total of $6 \times 34 = 204$ TDI combinations. Then considering the time reversal symmetry we get a total of 408. However, after those operations 198 combinations show to be trivial duplicates. Thus the total distinct TDI combinations are 210. 

\section{\label{svds} Single value decomposition to retrieve the acceleration noise signals}
Considering the single value decomposition, in our specific case we have $U_{i k}$ as a $92 \times 92$ matrix, $W_{kl}$ which is a $92 \times 54$ matrix, that has only the 7 uppermost diagonal elements different from zero, and $V_{l j}$ which is a $54 \times 54$ matrix. \\
Therefore the singular value decomposition of the acceleration matrix of coefficients $\delta_{ij}$ is
\begin{equation}
\delta_{ij} = \sum_{k=1}^{N}\sum_{l=1}^{54} U_{i k} W_{k l}V_{l j}^T, \label{eq:nullspaceeq}
\end{equation}
with $N=92$ the number of null combinations.
If we then multiply Eq. (\ref{eq:nullspaceeq}) by the inverse of $U_{i k}$ we get a $92 \times 54$ matrix that has only the first 7 lines different from zero:
\begin{equation}
\sum_{i=1}^{N} U_{m i }^{-1} \delta_{ij} = \sum_{l=1}^{54} W_{m l}V_{l j}^T.
\end{equation}
This allows us to redefine a new linear 92 combinations, still null to GW, as following: \begin{equation}
\tilde \psi_m =\sum_{i=1}^{92} \sum_{k=1}^{102} U_{m i }^{-1}\psi_k(t) =\sum_{i=1}^{92} \sum_{k=1}^{102} U_{m i }^{-1}  c_{ik} h[TDI_k],\label{eq:psimeq}
\end{equation}
for which only the first 7 entries have non-zero acceleration signal. $\psi_k(t)$ are the 102 TDIs that are a subset of the 210 channels (see Supplemental Material at [URL] for the full list). 
Thus only 7 have a physical meaning and, in this way, we have simplified the problem. \\
This reduced set of null-channels give us a better understanding of the reason why the rank R is 7. 
Indeed, only the signals for $m \leq R$ in Eq. (\ref{eq:psimeq}) are different from zero, as for any value of $m$ the GW part of the signal is zero. Besides for $m > R$ also the TM acceleration signal is zero.

\clearpage
\bibliography{bibliographyarticolo.bib}

\end{document}


Tables of TDI  combinations that suppress laser noise. $\text{N}_s[\#]$ is the sequence number. We have a total of 174 combinations of 16-links, 24 of 14-links and 12 of 12-links. capital $C$ are the core combinations listed in the Tables I, II and III, $c1$ and $c2$ are the correspondent two cyclical index permutations, $m1, m2$ and $m3$ are the three mirroring permutation of the core combination around $SC_1$, $SC_2$ and $SC_3$ respectively, and $tr-C$, $tr-c1$, $tr-c2$, $tr-m1, tr-m2$ and $tr-m3$ are the correspondent time reversal symmetries. Numbers within each sequence indicate the satellite on which each event takes place. First and last event coincide. Arrow indicate the direction of the link connecting adjoining events. Events at the start of two arrows represent simultaneous emission of two beams. Events at the end of two arrows represent measurements. The first SC is always an emission event. We can notice that not all the combinations have the entire set of the 6 satellites permutation symmetries but some of them show to have just the two cyclic permutations symmetries. The reason for this behaviour is due to the intrinsic symmetrical nature of some combinations. \\ For instance from TDI $C^{16}_1$, that is the well know X, by applying the two cyclic indexes permutations we get TDI Y and TDI Z. However we could obtain exactly the same output by mirroring $C_1$ around $SC_2$ and $SC_3$ and applying a string shift. In addition, $C^{16}_1$ is invariant under a time reversal of the sequence. Thus $C^{16}_1$ show to have just two symmetries out of seven.\\

\begin{table}[H]
\begin{center}
\resizebox{\textwidth}{.48\textheight}{
\begin{tabular}{| c | c | c || c || c | c |}
\hline
N$_s[\#]$ & channel & Sequence &  N$_s[\#]$  & channel  & Sequence  \\
\hline
N$_s[1]$ & $C^{16}_1$ & $ 1 \rightarrow 2 \rightarrow 1 \rightarrow 3 \rightarrow 1 \rightarrow 3 \rightarrow 1 \rightarrow 2 \rightarrow 1 \leftarrow 3 \leftarrow 1 \leftarrow 2 \leftarrow 1 \leftarrow 2 \leftarrow 1 \leftarrow 3 \leftarrow 1 $ & N$_s[88]$ & $C^{16}_{16}$  & $ 1 \rightarrow 2 \rightarrow 1 \rightarrow 3 \leftarrow 2 \leftarrow 1 \leftarrow 2 \rightarrow 3 \rightarrow 2 \leftarrow 1 \leftarrow 2 \leftarrow 3 \rightarrow 1 \rightarrow 2 \rightarrow 1 \leftarrow 3 \leftarrow 1 $   \\
N$_s[2]$ & \quad c1 & $ 2 \rightarrow 3 \rightarrow 2 \rightarrow 1 \rightarrow 2 \rightarrow 1 \rightarrow 2 \rightarrow 3 \rightarrow 2 \leftarrow 1 \leftarrow 2 \leftarrow 3 \leftarrow 2 \leftarrow 3 \leftarrow 2 \leftarrow 1 \leftarrow 2 $ & N$_s[89]$ & \quad c1 & $ 2 \rightarrow 3 \rightarrow 2 \rightarrow 1 \leftarrow 3 \leftarrow 2 \leftarrow 3 \rightarrow 1 \rightarrow 3 \leftarrow 2 \leftarrow 3 \leftarrow 1 \rightarrow 2 \rightarrow 3 \rightarrow 2 \leftarrow 1 \leftarrow 2 $   \\
N$_s[3]$ & \quad c2 & $ 3 \rightarrow 1 \rightarrow 3 \rightarrow 2 \rightarrow 3 \rightarrow 2 \rightarrow 3 \rightarrow 1 \rightarrow 3 \leftarrow 2 \leftarrow 3 \leftarrow 1 \leftarrow 3 \leftarrow 1 \leftarrow 3 \leftarrow 2 \leftarrow 3 $ & N$_s[90]$ & \quad c2 & $ 3 \rightarrow 1 \rightarrow 3 \rightarrow 2 \leftarrow 1 \leftarrow 3 \leftarrow 1 \rightarrow 2 \rightarrow 1 \leftarrow 3 \leftarrow 1 \leftarrow 2 \rightarrow 3 \rightarrow 1 \rightarrow 3 \leftarrow 2 \leftarrow 3 $   \\
\hline
   \end{tabular}}\caption{16-link TDI  combinations that suppress laser noise. The symbols are the same of Tables I}\label{T16all}
\end{center}
\end{table}

\begin{table}[H]
\begin{center}
\resizebox{\textwidth}{.05\textheight}{
\begin{tabular}{| c | c | c || c || c | c |}
\hline
N$_s[\#]$ & channel & Sequence &  N$_s[\#]$ & channel  & Sequence  \\
\hline
N$_s[175]$ & $C^{14}_{1}$ & $ 1 \rightarrow 2 \rightarrow 1 \rightarrow 3 \rightarrow 2 \rightarrow 1 \leftarrow 3 \leftarrow 2 \leftarrow 1 \leftarrow 2 \rightarrow 3 \rightarrow 1 \leftarrow 2 \leftarrow 3 \leftarrow 1 $ & N$_s[187]$ & $C^{14}_{2}$ & $ 1 \rightarrow 2 \rightarrow 1 \rightarrow 3 \leftarrow 2 \leftarrow 1 \leftarrow 3 \rightarrow 2 \leftarrow 1 \leftarrow 2 \leftarrow 3 \rightarrow 1 \rightarrow 2 \rightarrow 3 \leftarrow 1 $   \\
N$_s[176]$ & \quad c1 & $ 2 \rightarrow 3 \rightarrow 2 \rightarrow 1 \rightarrow 3 \rightarrow 2 \leftarrow 1 \leftarrow 3 \leftarrow 2 \leftarrow 3 \rightarrow 1 \rightarrow 2 \leftarrow 3 \leftarrow 1 \leftarrow 2 $ & N$_s[188]$ & \quad c1 & $ 2 \rightarrow 3 \rightarrow 2 \rightarrow 1 \leftarrow 3 \leftarrow 2 \leftarrow 1 \rightarrow 3 \leftarrow 2 \leftarrow 3 \leftarrow 1 \rightarrow 2 \rightarrow 3 \rightarrow 1 \leftarrow 2 $   \\
N$_s[177]$ & \quad c2 & $ 3 \rightarrow 1 \rightarrow 3 \rightarrow 2 \rightarrow 1 \rightarrow 3 \leftarrow 2 \leftarrow 1 \leftarrow 3 \leftarrow 1 \rightarrow 2 \rightarrow 3 \leftarrow 1 \leftarrow 2 \leftarrow 3 $ & N$_s[189]$ & \quad c2 & $ 3 \rightarrow 1 \rightarrow 3 \rightarrow 2 \leftarrow 1 \leftarrow 3 \leftarrow 2 \rightarrow 1 \leftarrow 3 \leftarrow 1 \leftarrow 2 \rightarrow 3 \rightarrow 1 \rightarrow 2 \leftarrow 3 $   \\
N$_s[178]$ & \quad m1 & $ 1 \rightarrow 3 \rightarrow 1 \rightarrow 2 \rightarrow 3 \rightarrow 1 \leftarrow 2 \leftarrow 3 \leftarrow 1 \leftarrow 3 \rightarrow 2 \rightarrow 1 \leftarrow 3 \leftarrow 2 \leftarrow 1 $ & N$_s[190]$ & \quad m1 & $ 1 \rightarrow 3 \rightarrow 1 \rightarrow 2 \leftarrow 3 \leftarrow 1 \leftarrow 2 \rightarrow 3 \leftarrow 1 \leftarrow 3 \leftarrow 2 \rightarrow 1 \rightarrow 3 \rightarrow 2 \leftarrow 1 $   \\
\hline
   \end{tabular}}
\end{center}
\caption{14-link TDI  combinations that suppress laser noise. The symbols are the same of Table I}
\label{T14all}
\end{table}

\begin{table}[H]
\begin{center}
\resizebox{\textwidth}{.04\textheight}{
\begin{tabular}{| c | c | c || c || c | c |}
\hline
N$_s[\#]$ & channel & Sequence &  N$_s[\#]$ & channel  & Sequence  \\
\hline
N$_s[199]$ & $C^{12}_{1}$ & $ 1 \rightarrow 2 \rightarrow 3 \rightarrow 1 \rightarrow 3 \rightarrow 2 \rightarrow 1 \leftarrow 3 \leftarrow 2 \leftarrow 1 \leftarrow 2 \leftarrow 3 \leftarrow 1 $ & N$_s[205]$ & \quad m1 & $ 1 \rightarrow 3 \rightarrow 2 \rightarrow 3 \rightarrow 1 \leftarrow 2 \leftarrow 3 \leftarrow 1 \rightarrow 2 \rightarrow 1 \leftarrow 3 \leftarrow 2 \leftarrow 1 $   \\
N$_s[200]$ & \quad c1 & $ 2 \rightarrow 3 \rightarrow 1 \rightarrow 2 \rightarrow 1 \rightarrow 3 \rightarrow 2 \leftarrow 1 \leftarrow 3 \leftarrow 2 \leftarrow 3 \leftarrow 1 \leftarrow 2 $ &N$_s[206]$ & \quad m2 & $ 3 \rightarrow 2 \rightarrow 1 \rightarrow 2 \rightarrow 3 \leftarrow 1 \leftarrow 2 \leftarrow 3 \rightarrow 1 \rightarrow 3 \leftarrow 2 \leftarrow 1 \leftarrow 3 $   \\
N$_s[201]$ & \quad c2 & $ 3 \rightarrow 1 \rightarrow 2 \rightarrow 3 \rightarrow 2 \rightarrow 1 \rightarrow 3 \leftarrow 2 \leftarrow 1 \leftarrow 3 \leftarrow 1 \leftarrow 2 \leftarrow 3 $ & N$_s[207]$ & \quad m3 & $ 2 \rightarrow 1 \rightarrow 3 \rightarrow 1 \rightarrow 2 \leftarrow 3 \leftarrow 1 \leftarrow 2 \rightarrow 3 \rightarrow 2 \leftarrow 1 \leftarrow 3 \leftarrow 2 $   \\
N$_s[202]$ & $C^{2}_{2}$ & $ 1 \rightarrow 2 \rightarrow 3 \rightarrow 2 \rightarrow 1 \leftarrow 3 \leftarrow 2 \leftarrow 1 \rightarrow 3 \rightarrow 1 \leftarrow 2 \leftarrow 3 \leftarrow 1 $ & N$_s[208]$ & $C^{12}_{3}$ & $ 1 \rightarrow 2 \rightarrow 1 \leftarrow 3 \rightarrow 2 \leftarrow 1 \rightarrow 3 \leftarrow 2 \leftarrow 3 \rightarrow 1 \leftarrow 2 \rightarrow 3 \leftarrow 1 $   \\
\hline
 \end{tabular}}
\end{center}
\caption{12-link TDI combinations that suppress laser noise. The symbols are the same of Table I}
\label{T12all}
\end{table}
Linear combinations of the 210 laser noise cancelling TDI  whose signals, for normal wave propagation, have zero GW component and non zero TM motion component
\begin{table}
\centering
 \caption{\label{combinations} \footnotesize{ TDI combinations whose signals, for normal wave propagation, have zero GW component and non zero TM motion component}}
\resizebox{\textwidth}{.48\textheight}{
\begin{tabular}{|c|c|c|c|c|c|} \hline
$\#$ & Combinations & $\#$ & Combinations & \# & Combinations \\  \hline
 1 & Ns(210) & 60 & -Ns(19)-Ns(20)+Ns(25)+Ns(26)+Ns(32)+Ns(139) & 119 & Ns(19)-Ns(25)-Ns(31)-Ns(32)+Ns(76) \\  \hline
 2 & Ns(209) & 61 & Ns(19)+Ns(20)-Ns(25)-Ns(26)+Ns(138) & 120 & Ns(19)+Ns(20)-Ns(25)-Ns(26)-Ns(32)+Ns(75) \\  \hline
 3 & Ns(208) & 62 & -Ns(19)-Ns(20)+Ns(25)+Ns(26)+Ns(135) & 121 & -Ns(20)+Ns(26)-Ns(31)+Ns(74) \\  \hline
 4 & Ns(20)-Ns(26)+Ns(31)+Ns(207) & 63 & Ns(20)-Ns(26)+Ns(134) & 122 & -Ns(19)+Ns(25)+Ns(31)+Ns(32)+Ns(73) \\  \hline
 5 & -Ns(19)-Ns(20)+Ns(25)+Ns(26)+Ns(32)+Ns(206) & 64 & Ns(19)-Ns(25)+Ns(133) & 123 & -Ns(19)+Ns(25)+Ns(26)+Ns(31)+Ns(32)+Ns(72) \\  \hline
 6 & Ns(19)-Ns(25)-Ns(31)-Ns(32)+Ns(205) & 65 & -Ns(19)-Ns(20)+Ns(25)+Ns(26)+Ns(32)+Ns(132) & 124 & -Ns(20)-Ns(25)-Ns(31)+Ns(71) \\  \hline
 7 & Ns(19)+Ns(20)-Ns(25)-Ns(26)-Ns(32)+Ns(204) & 66 & Ns(19)-Ns(25)-Ns(31)-Ns(32)+Ns(131) & 125 & Ns(19)+Ns(20)-Ns(26)-Ns(32)+Ns(70) \\
\hline
\end{tabular}}
\end{table}